%% file: ArXiv_PVC_Thermoacoustics.tex
\documentclass[a4paper,10pt]{article}
\usepackage{geometry}
\usepackage{graphicx}
\usepackage{epstopdf, epsfig}
\usepackage{xcolor}
\usepackage{mathrsfs}

\usepackage{amssymb,amsmath}
\usepackage{tabto}
\usepackage{relsize}
\usepackage{textcomp}

\usepackage[utf8]{inputenc}
\usepackage{natbib}

\usepackage{fancyhdr}
\title{Impact of precessing vortex core dynamics on the thermoacoustic instabilities in a swirl stabilized combustor}

\usepackage{authblk}
\author[1]{Ashwini Karmarkar}
\author[2]{Saarthak Gupta}
\author[3]{Isaac Boxx}
\author[1]{Santosh Hemchandra}
\author[2]{Jacqueline O'Connor}

\affil[1]{Department of Mechanical Engineering, Pennsylvania State University, University Park, PA}
\affil[2]{Department of Aerospace Engineering, Indian Institute of Science, Bengaluru, India}
\affil[3]{DLR - German Aerospace Center, Stuttgart, Germany}

\date{}
\begin{document}

\maketitle

\begin{abstract}
Global instabilities in swirling flows can significantly alter the flame and flow dynamics of swirl-stabilized flames, such as those in modern power generation gas turbine engines. In this study, we characterize the interaction between the precessing vortex core (PVC), which is the consequence of a global hydrodynamic instability, and thermoacoustic instabilities, which are the result of a resonant coupling between combustor acoustics and the unsteady heat release rate of combustion. This characterization is performed using experimental data obtained from a model gas turbine combustor system employing two concentric swirling nozzles of air, separated by a ring of fuel injectors, operating at 5 bar pressure. The flow split between the two streams is systematically varied to observe the impact of flow structure variation on the flow and flame dynamics. High-speed stereoscopic particle image velocimetry, OH planar laser-induced fluorescence, and acetone planar laser-induced fluorescence are used to obtain information about the velocity fields, flame, and fuel flow behavior, respectively. Spectral proper orthogonal decomposition and spatial frequency analysis are used to identify and characterize the dominant oscillation mechanisms driving the system. Three dominant modes are seen: two thermoacoustic modes and the precessing vortex core. Our results show that in the cases where the frequency of the PVC overlaps with either of the thermoacoustic modes, the thermoacoustic modes are suppressed. A weakly nonlinear asymptotic analysis shows that the suppression of the axisymmetric shear layer shedding, and hence thermoacoustic mode, is the result of a nonlinear coupling between the PVC and the axisymmetric mode of the swirling jet. Evolution equations for both the symmetric and PVC modes are derived to show the controlling parameters that drive this suppression. We conclude by discussing ways in which thermoacoustic instability suppression can be achieved through combustor flow field design.
\end{abstract}
\section{Introduction}
For modern power generation gas turbines, increasingly stringent requirements on $NO_x$ emissions have driven a technology shift towards lean premixed burners. While fuel lean, premixed systems allow combustors to achieve significantly lower $NO_x$ emissions, these flames are inherently less stable and hence more susceptible to acoustic disturbances, which can be a precursor to the occurrence of combustion instabilities \citep{correa1998gasturbines,lefebvre98GTbook}. Combustion instability is a thermoacoustic instability that arises from a coupling between the resonant acoustic modes of the combustor and the unsteady heat release rate of the flame. Occurrence of combustion instability is a limiting factor to engine performance and operability, and, in extreme cases, can lead to severe damage of engine hardware \citep{lieuwen_yang2005combustion_instability,candel2002cdc,correa1998gasturbines}.
\par
The acoustic modes of a combustor can couple with the unsteady rate of heat release through fluctuations in pressure, velocity, and/or equivalence ratio \citep{lieuwen2012unsteady}. Velocity coupling mechanisms can be driven by large-scale coherent structures in the flow that distort the flame and consequently alter the heat release rate \citep{paschereit1999coherent, santavicca2005flamevortex_instability, poinsot1987vortex,lieuwen2009bluffbody_flames_forcing,ratner2007acoustic_forcing_flame}. For instance, it has been shown that vortices in the flow can entrain fresh reactants and cause the flame to roll-up, which sharply increases the flame surface area, thereby creating a heat release pulse. This coherent oscillation in heat release feeds back into the acoustic pressure fluctuations and creates a self-sustaining thermoacoustic instability \citep{candel1999flamevortex,santavicca1993flamevortex,ghoniem1988flamevortex,poinsot1987vortex}. A detailed review of the different mechanisms by which flames interact with vortical structures is provided by \cite{renard2000flamevortex}. 
\par 
Equivalence ratio fluctuations have also been shown to be drivers of combustion instabilities in lean premixed combustors \citep{lieuwen1998ER_coupling,santavicca2010er_vcoupling,shreekrishna2010er_coupling,sattelmayer2006er_fluctuations,Oberleithner2019ER_coupling}. Pressure oscillations in the combustor interact with the fuel supply and cause coherent oscillations in the equivalence ratio of the reactant mixture, which when convected to the flame, can cause pulsations in the heat release rate. \cite{santavicca2000ER_measurement} have shown, using an infrared absorption technique to measure equivalence ratio fluctuations, that the equivalence ratio fluctuations are strongly linked to the heat release fluctuations in an unstable combustor and can thereby play a significant role in driving the instability.
\par
In partially-premixed, swirl-stabilized systems used in gas turbine combustors, such as the combustor considered in this work \citep{Stohr20195421}, complex hydrodynamic and thermo-chemical processes are involved, which may lead to multiple coupling pathways being present in the system. The dynamics of swirling flows are especially complex because swirling flows can exhibit several modes of hydrodynamic instability. Swirling flows are typically characterized by a non-dimensional parameter known as the swirl number, which is the ratio of the axial flux of tangential momentum to the axial flux of axial momentum \citep{gupta84swirl}. Strongly swirling flows, such as those seen in gas turbine combustors, are characterized by the formation of a central recirculation zone, or a `vortex breakdown bubble' \citep{harvey62VB,hall72VBreview}. This recirculation zone constantly supplies hot products to the base of the flame, thereby enhancing its static stability. 
\par
Increasing the degree of swirl can cause this recirculation zone to precess about the axis of symmetry, forming what is known as the \textit{precessing vortex core} (PVC) \citep{syred06review}. PVCs are the the manifestation of a global instability obtained through a super-critical Hopf bifurcation induced by vortex breakdown \citep{manoharan2020PVC,oberleithner2011stabilityanalysis}. \cite{manoharan2020PVC} have shown, using experimental data and results from a weakly nonlinear stability analysis, that beyond a critical swirl number, the frequency and square of the amplitude of the PVC oscillations scale linearly with swirl number. They have also confirmed from ab initio estimates of the coefficients of the associated Stuart-Landau equation for the oscillation amplitude of the PVC that it is a stable limit cycle flow oscillation that results from a supercritical Hopf bifurcation in the flow state.
\par
In combustor systems, the occurrence of a PVC is a function of swirl number, flame shape, fuel/air mixing, combustor configuration, and equivalence ratio \citep{syred06review}. PVC dynamics can significantly influence the flame and flow dynamics in a number of ways \citep{candel2014swirl_review}. The shear layer oscillations due to the perturbations caused by the precession of the vortex breakdown bubble have been shown to interact with the flame and cause flame surface distortion \citep{stohr2012vortexflame,stohr2015mixing}. The occurrence of PVCs can also, in some cases, impact the stability and structure of swirl stabilized flames \citep{oberandstohr15pvc}. It has also been shown that PVCs can enhance fuel/air mixing \citep{freitag05pvc,stohr2015mixing}.  
\par
The primary focus of this study is to characterize the impact of the PVC on the thermoacoustic modes of the combustor in a partially-premixed flame at elevated pressure. The interaction of hydrodynamic instabilities, such as the PVC and symmetric acoustic modes, is of significant interest as both these modes can be simultaneously observed in many combustor configurations. \cite{hemchandra2018hydrodynamics} proposed two possible coupling pathways between self-excited hydrodynamic instability modes and the combustor acoustic modes that can lead to combustion instability. These results indicate that the strength of the coupling between the hydrodynamic and acoustic modes depends on how close the instability frequency is to the combustor acoustic eigenfrequency, in addition to how effectively acoustic modes can excite the flow in the regions of hydrodynamic mode receptivity. 
\par
\cite{steinberg2010PVC} showed that in a case where both the PVC and thermoacoustic mode are simultaneously present, the PVC can undergo axial extension and contraction at the thermoacoustic frequency. \cite{moeck2012PVCandAcoustics} and \cite{steinberg2010PVC} reported the presence of a spectral peak at an interaction frequency that corresponds to the difference between the PVC and thermoacoustic frequencies and suggested that this interaction frequency is likely a consequence of nonlinear phenomena. \cite{oberleithner2019pvc_acoustics} studied the impact of actuation of helical flow modes using a centerbody on their interaction with thermoacoustic modes. Their results showed that, in the case of a partially-premixed flame, this actuation can significantly reduce thermoacoustic instability. They suggested that this suppression of thermoacoustic instability modes is likely a consequence of suppression of equivalence ratio fluctuation due to enhanced mixing caused by the helical flow mode excited. By contrast, in some cases, the self-excited PVC mode is suppressed in reacting flows \citep{oberandstohr15pvc,oberleithner2013pvc_densitygradient}. \cite{taamallah2016} showed that the presence of a PVC in a premixed swirl-stabilized combustor can provide a low velocity path within the combustor flow field resulting in a flame shape change. This new flame shape, in turn, causes the combustor acoustic mode to couple with the flame, resulting in thermoacoustic oscillations.
\par
Previous work from some of the present authors showed, through a spectral analysis of a swirling flow, that response of the shear layers to acoustic forcing in an isothermal swirling jet can be suppressed in the presence of a PVC \citep{mathews2016PVC}. This result was explained by means of a linear stability analysis, showing that the formation of the PVC results in increased shear layer thickness. This increased thickness leads to a progressive weakening of the receptivity of the axisymmetric Kelvin-Helmholtz mechanism to imposed forcing \citep{oconnor18PVCshear}. 
\par 
Put together, the existing literature shows that the dynamics of the PVC can have a critical impact on the performance of swirl-stabilized combustion systems. Furthermore, the PVC can interact with the symmetric acoustic modes present in the combustor in a number of ways, depending on the operating conditions and combustor configuration. This study extends these concepts to a more realistic configuration with a dual-annular swirl injector, commonly found in gas turbine combustors \citep{joshi1998dry}, partially-premixed fuel injection, and elevated pressures. Using an optically accessible model swirl-stabilized gas turbine combustor configuration, operating at an elevated pressure of 5 bar, we study the interaction between the dominant oscillation modes present in the system and observe the evolution of this interaction at a range of flow conditions. Our results show that, in the configuration studied, two dominant thermoacoustic modes are present. In cases where all three modes are simultaneously present, the PVC oscillation mode is generally the governing mode in the flow field. Furthermore, whenever the frequency of the PVC approaches that of either thermoacoustic mode, the thermoacoustic mode is subsequently suppressed. 
\par
In order to understand the interaction between the PVC and the thermoacoustic modes, we extend the weakly nonlinear analysis of \citep{manoharan2020PVC} to variable density flows and include a stable hydrodynamic axisymmetric mode weakly forced by thermoacoustic oscillations near its natural frequency. Similar analysis has been performed recently by \cite{rigas2017} in the context of axisymmteric wakes. We obtain a closed-form expression for the steady-state amplitude of the axisymmetric mode oscillating at the thermoacoustic oscillation frequency. This steady-state amplitude has a strong dependence on the difference between the thermoacoustic and the natural PVC frequencies. The outcome of this dependence is that as the PVC frequency approaches the thermoacoustic oscillation frequency, it causes the suppression of the axisymmetric hydrodynamic, which can then result in the suppression of the thermoacoustic oscillation. The analysis also shows that the nonlinear coupling between the PVC and the axisymmetric hydrodynamic mode results in helical oscillations at the interaction frequencies, i.e., the sum and difference of the thermoacoustic and PVC oscillation frequencies. We restrict our theoretical analysis in the present paper to a qualitative description of the interaction between the PVC and the axisymmetric modes. The time-averaged velocity and density fields are required to quantify the various contributions to the steady-state solution. While it would be interesting to analyze various contributions and the influence of nonlinear coupling, the spatial distribution of the time-averaged density field, which has a significant impact on the hydrodynamic modes \citep{kiran15instability, emerson2012density, kiran2014bfsc}, is difficult to obtain from experiments of the present kind with sufficient quantitative accuracy. Therefore, quantitative confirmation of the theoretical analysis is beyond the scope of this paper.

\section{Experimental Methods}
\subsection{Burner and operating conditions}
\par
\begin{figure}
    \centering
    \includegraphics[width=70mm]{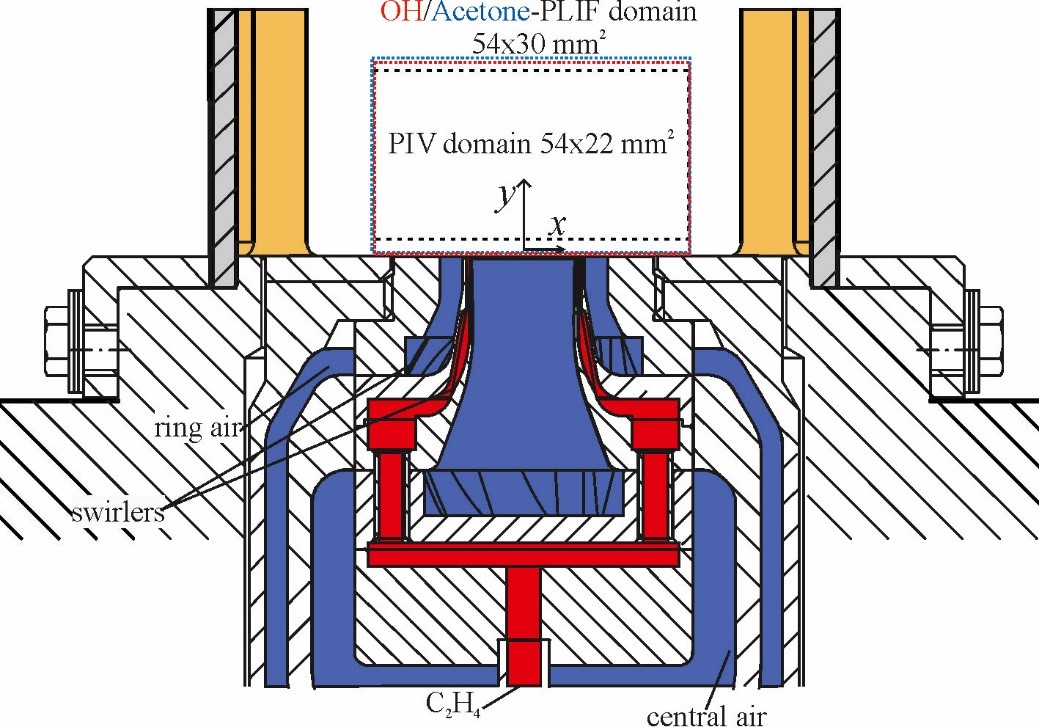}
    \caption{Experimental Setup}
    \label{fig:dlr_experiment}
\end{figure}

The burner used in the current study is shown in Fig. \ref{fig:dlr_experiment}; its geometry has been described in previous publications \citep{geigle2015dlr_burner, geigle2017flow}. For completeness, it is briefly described again here. The injector consists of a pair of annular air swirl nozzles separated by a ring of 60 fuel channels, each with an area of 0.5$\times$0.4 $mm^2$, for the introduction of gaseous fuel, ethylene. The central air nozzle has a diameter of 12.3 mm. The annular outer air nozzle measures 19.8 mm in diameter. The combustion chamber has a square cross section measuring 68$\times$68 $mm^2$. It is 120 $mm$ long and enclosed by 3-mm-thick quartz windows for optical access. The water-cooled dome of the combustion chamber has a cylindrical exhaust hole - diameter 40 mm, length 24 $mm$ - linked to the combustion chamber by a curved tube. The combustor was mounted in an optically accessible pressure vessel.
    
The flows of air (central, annular and seeding) and fuel were supplied by separate mass ﬂow controllers (Bronkhorst). The flow rate of air is 841.7 SLM and flow rate of fuel is 37.9 SLM, resulting in a thermal power of 38.4 kW. The flow controllers were calibrated in-house, resulting in an accuracy of better than 1.0\% of the controllers’ maximum ﬂow. The majority of the ethylene flow was bubbled through a temperature-controlled reservoir (T=$25^{\circ}$C, 298 K) filled with acetone, which provided an acetone-saturated ethylene flow of the desired composition. For the present study, the combustor was operated with ethylene/acetone fuel at a pressure of 5 bars and $\phi = 0.67$. The air split, defined as the fraction of total air flow passing through the center nozzle, was varied from 0.2 -- 0.5. The Reynolds number based on the diameter of the outer nozzle is $19,490$.

\subsection{Measurement techniques}
Time-resolved stereo particle image velocimetry (sPIV), planar laser-induced fluorescence of OH (OH-PLIF), and acetone-PLIF were applied simultaneously with a repetition rate of 10 kHz to capture flame/vortex/fuel interaction. Run duration was 1 s in the technically premixed ethylene/air swirl flame stabilized in the model of the aero-engine combustor. This diagnostic system has been described previously in the scientific literature \citep{Litvinov2020DLR}.  For completeness, a brief description is included below. 

\subsubsection{Particle Image Velocimetry}
Velocity fields were measured using a two-camera, stereoscopic particle image velocimetry system. The sPIV system used a dual-cavity, diode-pumped, solid state laser (Edgewave, IS200-2-LD, up to 9 mJ/pulse at 532 nm) and a pair of highspeed CMOS cameras (Phantom V1212). The cameras were mounted on opposite sides of the laser sheet, looking down into the combustor. PIV image pairs were acquired at 10 kHz and 640$\times$ 800 pixel resolution. The PIV measurement domain spans across most of the width of the combustion chamber ($-27 mm < x < 27 mm$). In the axial direction, the domain spans $5 mm < y < 27 mm$,  as shown in Fig. \ref{fig:dlr_experiment}. Time separation of the laser pulses was $\delta t = 10 \mu s$. The beam was formed into a sheet using a pair of cylindrical lenses ($f = -38$ mm and 250 mm) and thinned to a waist using a third cylindrical lens ($f = 700$ mm). Both combustion air flows were seeded with titanium dioxide ($TiO_2$) particles of nominal diameter 0.5 $\mu m$. Image mapping, calibration, and particle cross-correlations were completed using a commercial, multi-pass adaptive window offset cross-correlation algorithm (LaVision DaVis 8.4). Final interrogation window size and overlap were 24x24 pixels and 50\%, respectively, for a spatial resolution of 1.9 mm and vector spacing of 0.95 mm. The absolute values of uncertainty of instantaneous velocities based on the correlation statistics in DaVis were estimated to be 0.1 $m/s$---1.1 $m/s$ for the in-plane components and 0.3 $m/s$---2.3 $m/s$ for the out-of-plane component. On average, uncertainty values for the in-plane components are in the range of 0.04 $m/s$---0.45 $m/s$ and the uncertainty values for the out-of-plane component are in the range of 0.10 $m/s$---0.9 $m/s$.

\subsubsection{OH- and acetone-PLIF}
The OH-/acetone-PLIF imaging system is based on a frequency-doubled dye laser, pumped by high-speed, pulsed Nd:YAG laser (Edgewave IS400-2-L, 135 W at 532 nm and 10kHz) and a pair of intensified high-speed CMOS camera systems. The dye laser system (Sirah Credo) produced 5.3 – 5.5 W at 283 nm and 10 kHz repetition rate (0.53-0.55 mJ/pulse). The dye laser was tuned to excite the Q1(9) and Q2(8) lines of the $A^2 \Sigma^+ - X^2 \Pi (v'=1,v''=0)$ band. These transitions merge at high pressure due to increased collisional line broadening, which mitigates to some degree fluorescence signal loss due to collisional line broadening. The laser wavelength was monitored continuously throughout the experiments using a photomultiplier tube (PMT) mounted to a 10 cm monochrometer and a premixed, laminar reference flame. The 283 nm PLIF excitation beam is formed into a sheet approximately 40 mm (high) x 0.2 mm (thick) using three fused-silica, cylindrical lenses (all anti-reflective coated to maximize transmission). The laser sheets of the OH-/acetone-PLIF and PIV systems were overlapped by passing the (green, 532 nm) PIV sheet through the final OH-PLIF turning mirror. 
\par
The same laser was used to excite fluorescence signal in both the OH- and the acetone-PLIF imaging systems. For acetone-PLIF, fluorescence signal was imaged via a CMOS camera (LaVision HSS8), an external two-stage intensifier (LaVision HS-IRO), equipped with 85 mm focal length, f/1.2 (Canon) objective and a band-pass interference filter. The filter (LOT, 450FS40-50) was centered at 450 nm for detection of the acetone fluorescence and had a bandpass of $\pm 20$ nm. OH-PLIF fluorescence signal was imaged using a similar highspeed CMOS camera (LaVision HSS8) and external two-stage intensifier (LaVision HS-IRO) from the opposite side of the combustor. The OH-PLIF camera was equipped with 64 mm focal length, f/2 (Halle) UV-objective and a high transmission bandpass interference filter. The OH-/acetone-PLIF measurement domain was slightly larger than that of the PIV system, spanning ($-27 mm< x < 27 mm$, $0 mm< y < 30 mm$), as shown in Fig. \ref{fig:dlr_experiment}. 
\par
Previous studies have demonstrated the feasibility of using acetone as fuel tracer in a dual-swirl GT model combustor similar to that used in the present study \citep{stohr2015mixing, stohr2017interaction}. These studies indicated that replacement of 10\% vol of the original (methane) fuel by acetone vapor did not significantly change the flame shape or dynamics. Although the present study uses ethylene rather than methane, it is expected to have a similarly negligible effect on the flame shape and dynamics. Data from the acetone PLIF signal is only considered until downstream distances of ~$z=15$ mm, as the fuel is largely consumed by that point. Signal downstream of that point is likely the result of fluorescence of polycyclic aromatic hydrocarbons or even weak soot luminescence, and so is not analyzed in the frequency analysis discussed in Sec. \ref{sec:flame}.

\subsection{Data analysis}
In this study, we predominantly use spectral proper orthogonal decomposition (SPOD) to characterize the frequency domain behavior of the flame and flow field. Proper orthogonal decomposition (POD) \citep{lumley1993pod} is an eigenvalue decomposition that yields energy-ordered modes that can be used to extract coherent structures from a flow field. Spectral POD is derived from a space-time POD problem for statistically stationary flows and leads to modes that each oscillate at a single frequency. The SPOD method used here is taken from \cite{towne2018spectral}. We use SPOD to extract high-energy coherent structures rather than using a traditional snapshot POD method, which provides spatially-orthogonal, energy-ordered modes of the data, because POD does not discern motions based on their spatiotemporal evolution, just their spatial correlations. As a result, tonal but spatially uncorrelated or low-energy oscillations are not necessarily separated from other motions in the POD, but can be identified in the SPOD. SPOD achieves this by calculating the eigenvalue problem on the cross-spectral density tensor, which results in spatial modes analogous to this from snapshot POD, but also provides frequency-resolved energy information about each of these modes. \cite{towne2018spectral} rigorously show that SPOD is an optimized dynamic mode decomposition (DMD) \citep{schmid_2010} method for stationary flows.

\section{Experimental Results}
\subsection{Time-averaged flow and flame profiles}
\begin{figure}
    \centering
    \includegraphics[width=100mm]{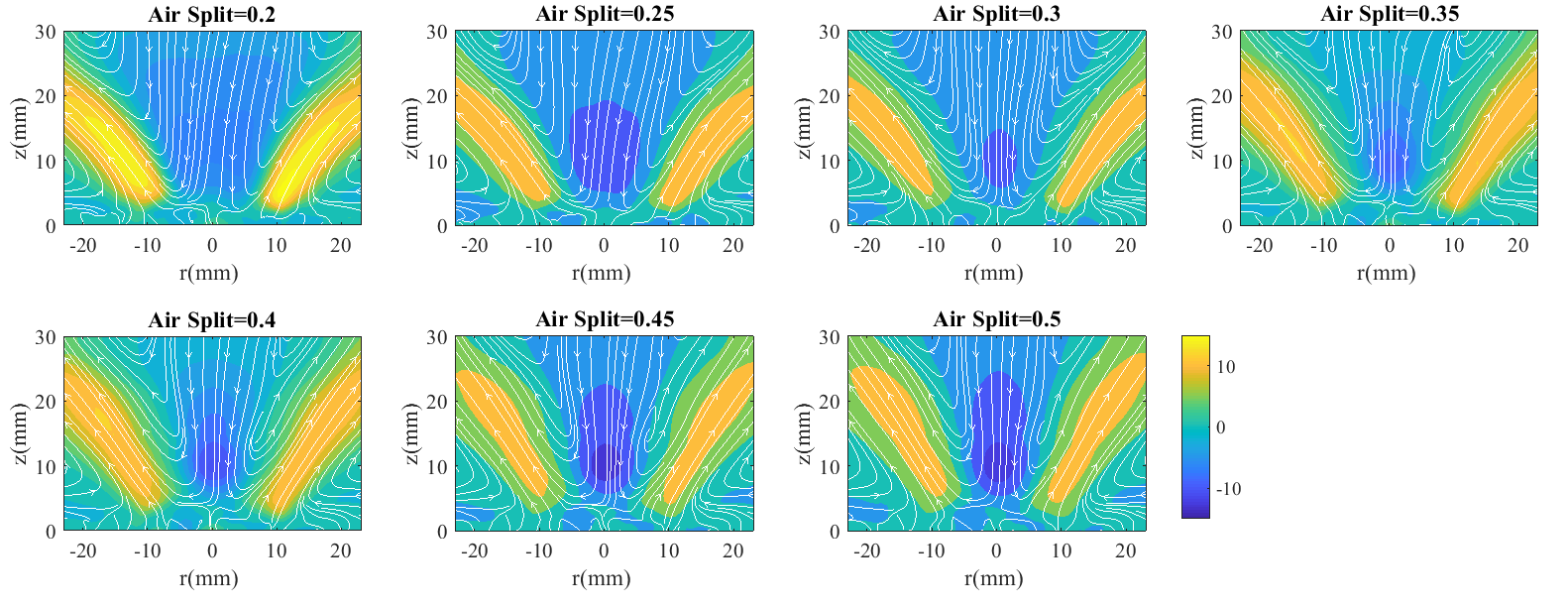}
    \caption{Time-averaged stream-wise velocity component with streamlines.}
    \label{fig:dlr_uavg}
\end{figure}

The time-averaged streamwise velocity contours with streamlines, obtained from the sPIV measurement, for all air split conditions are shown in Fig. \ref{fig:dlr_uavg}. In all cases, a region of negative axial velocity, the central recirculation zone, can be seen close to the axis of symmetry. As air split is increased, this recirculation becomes stronger and more well defined. 

\begin{figure}
    \centering
    \includegraphics[width=100mm]{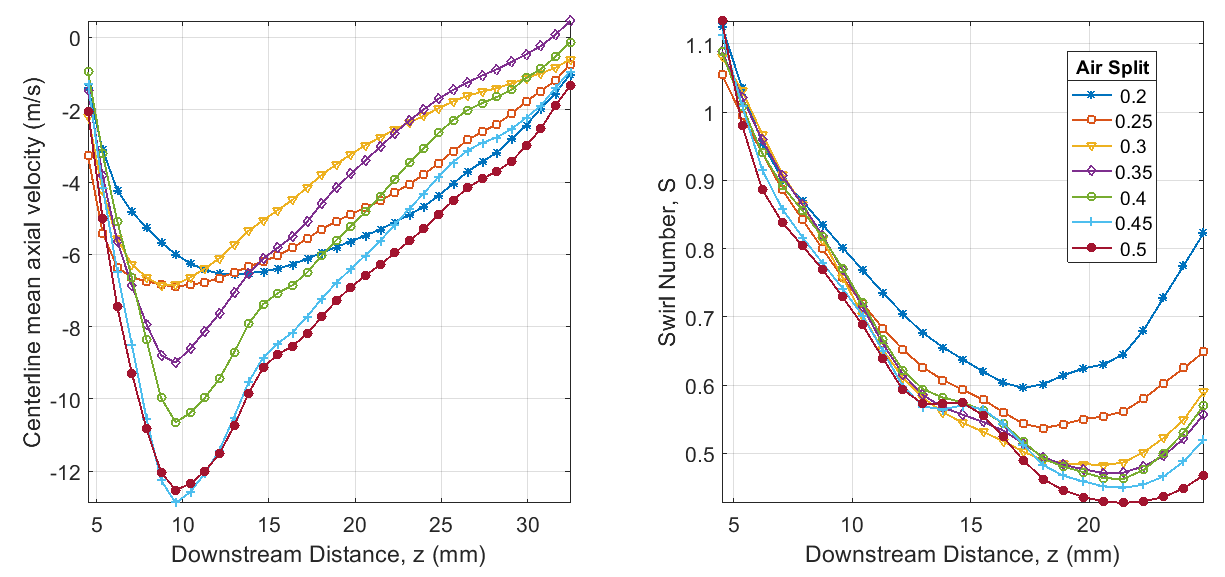}
    \caption{Centerline time-averaged axial velocity (left) and swirl number (right) as a function of downstream distance for every air split condition.}
    \label{fig:min_u}
\end{figure}

Fig. \ref{fig:min_u} shows the variation of the time-averaged axial velocity along the centerline and swirl number, derived from experimental data, $S_d$, as defined in eq. \ref{eq:swirlEq}, with downstream distance at every air split condition. To calculate $S_d$, the axial flux of azimuthal velocity is radially integrated from the $u_z=0$ contour at the outer boundary of the recirculation zone, at $r=r_c$, to the edge of the frame of data at $r=R$, at each downstream distance.

\begin{equation}
    S_d=\frac{\int_{r_c}^{R} u_\theta(z,r) u_z(z,r) r dr}{\int_{r_c}^{R} u_z(z,r)^2 r dr}
    \label{eq:swirlEq}
\end{equation}

The plots show that as air split increases, the strength of recirculation increases since the minimum velocity inside the vortex breakdown bubble systematically decreases. Furthermore, the streamwise position of the recirculation zone stays largely constant with varying air split. However, the swirl number is relatively constant for all the cases, which indicates that the change in the recirculation zone strength is a result of the change in the critical swirl number for vortex breakdown with air split, rather than a change in swirl number itself; this result is discussed further in Section \ref{sec:theory}, as is the impact of the strength and location of the recirculation zone on the hydrodynamic stability of the flow.

\begin{figure}
    \centering
    \includegraphics[width=100mm]{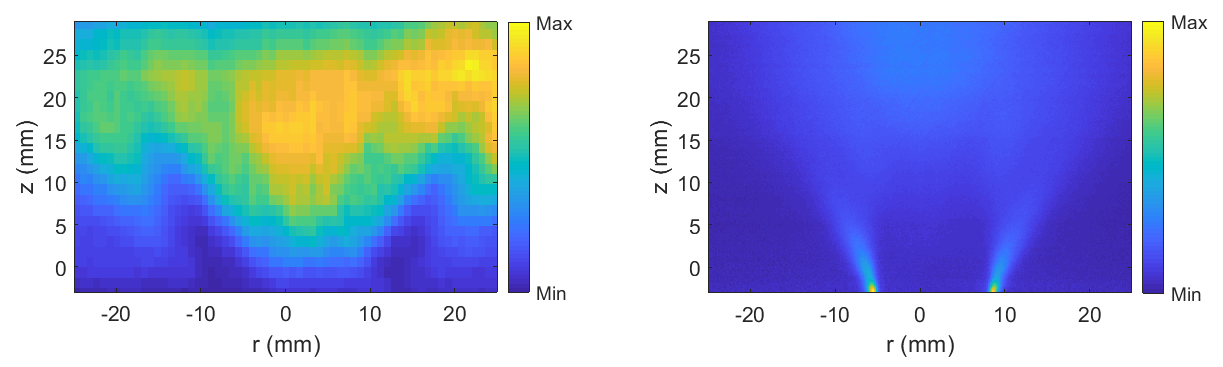}
    \caption{Time-averaged flame structure (left) and time-averaged fuel distribution profile (right) for air split=0.4 case}
    \label{fig:dlr_oh_acetone_avg}
\end{figure}

The time-averaged flame structure (obtained from time-averaging the OH-PLIF images) is shown in Fig. \ref{fig:dlr_oh_acetone_avg} for an air split of 0.4. Since this is a partially-premixed flame, and the operating pressure is relatively high, the OH-PLIF has a low signal-to-noise ratio and so the instantaneous flame shape is not always apparent The M-flame shape is evident from the time-averaged image in Fig. \ref{fig:dlr_oh_acetone_avg}, but the figure should not be interpreted quantitatively. The time-averaged fuel distribution (obtained from acetone-PLIF measurements) is shown also in Fig. \ref{fig:dlr_oh_acetone_avg} at the same air split of 0.35. The time-averaged profiles for the flame and fuel distribution do not change much with air split and so only one condition is provided here. In this case, as in all cases, it's evident that the flame has an M-flame shape that is fueled by the acetone jet located between the two swirl passages.

\par

\subsection{Spectral characterization of flow fields}
The time-averaged flow profiles suggest that increasing air split leads to strengthening of recirculation in the vortex breakdown region. This increase in recirculation strength can lead to changes in the hydrodynamic instability characteristics of the flow. To gain more insight into these dynamics of the flow, and to identify the governing oscillation modes in the system, the frequency domain characteristics are analyzed.
\par
\begin{figure}
    \centering
    \includegraphics[width=100mm]{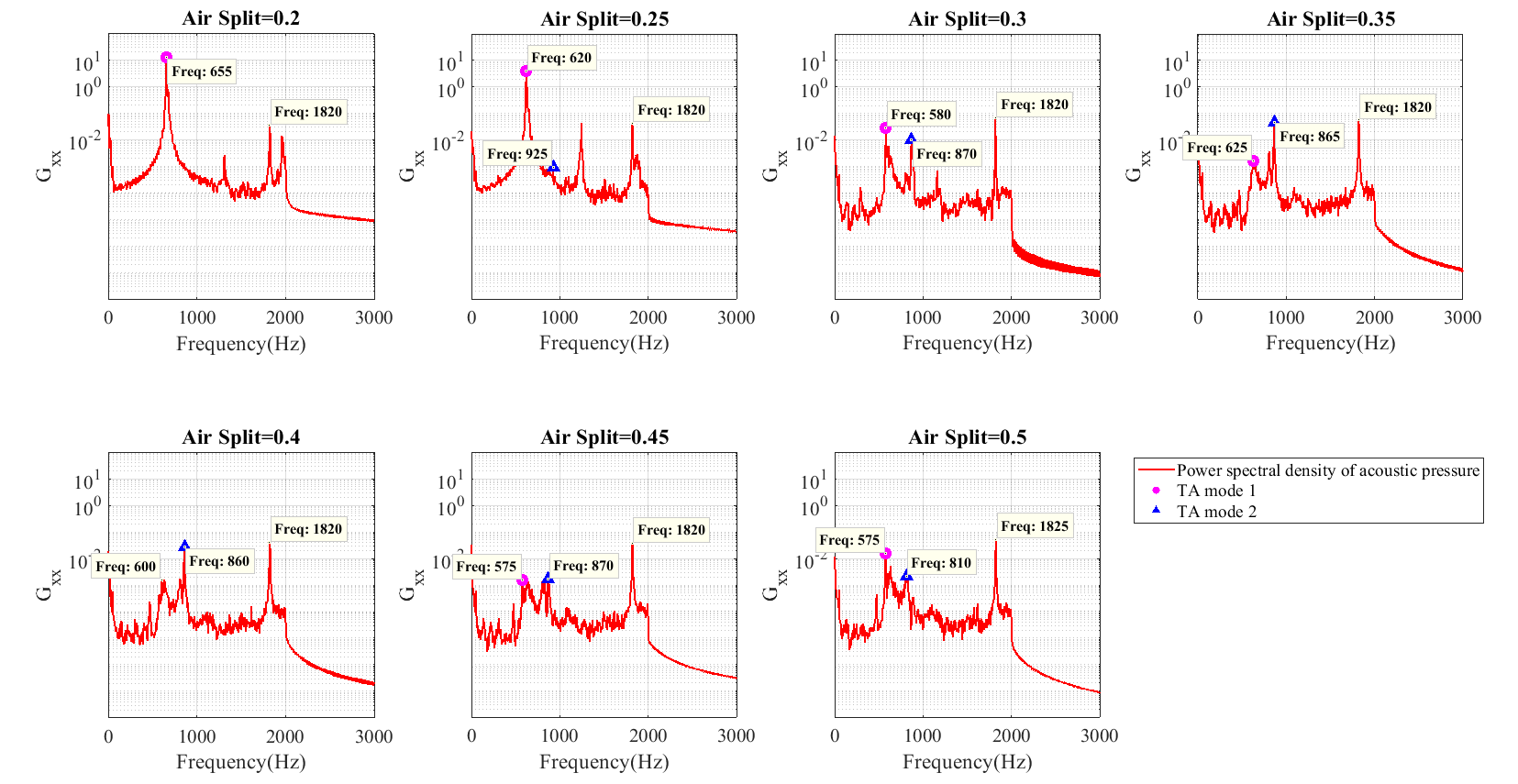}
    \caption{Acoustic pressure spectra at each air split condition.}
    \label{fig:dlr_acoustic}
\end{figure}

The combustor acoustic pressure spectra, obtained by performing an ensemble-averaged Fourier transform and calculating the power spectral density of the pressure data at each air split condition, is shown in Fig. \ref{fig:dlr_acoustic}. Pressure is measured near the outlet of the combustor. The power spectral density shows multiple distinctive peaks at each condition, indicating the presence of multiple sources of oscillation, which is characteristic of reacting systems exhibiting hydrodynamic and thermoacoustic instability. In the lower air-split cases, a strong peak can be seen at around 600 Hz, which is identified to be the thermoacoustic instability, marked by a circle. As air split increases, the peak at around 600 Hz persists, but is not as strong, and new peaks can be seen at around 800 Hz, which is later identified as as second thermoacoustic mode, marked with a triangle. Furthermore, harmonics of the dominant peaks can also be seen. In order to identify coherent modes in the flow, the spatial profiles of the flow fields at each of these modes are calculated using SPOD. 

\begin{figure}
    \centering
    \includegraphics[width=100mm]{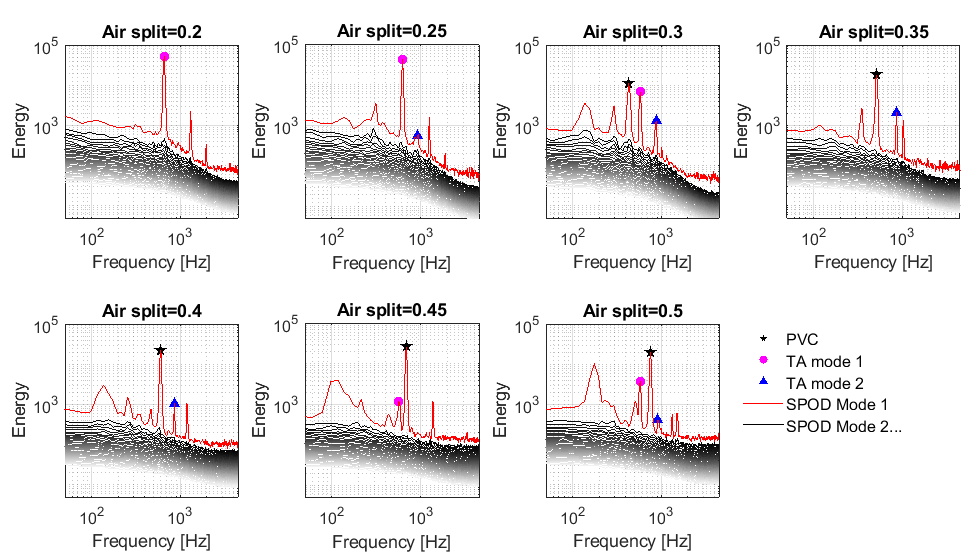}
    \caption{Eigenvalue spectra of velocity field obtained from SPOD.}
    \label{fig:spod_velocity}
\end{figure}

In this study, we use spectral proper orthogonal decomposition \citep{towne2018spectral} to characterize the frequency domain behavior of the flow and to extract the high energy coherent motion in the flow field. The eigenvalue spectra of the velocity data at every air split condition is shown in Fig. \ref{fig:spod_velocity}. The red curve in each plot depicts the highest energy mode and the gray curves depict subsequent modes in order of decreasing energy. In all cases, most of the coherent content is captured by the first mode, which is why the mode shapes of the first mode are analyzed. Similar to the acoustic pressure spectra, multiple peaks can be seen at each condition, which indicates that there are multiple coherent oscillations in the system. In order to understand the nature of the oscillation corresponding to every peak, the spatial mode shapes of the first mode at the respective peak frequencies are reconstructed at different operating conditions.

\par
\begin{figure}
    \centering
    \includegraphics[width=75mm]{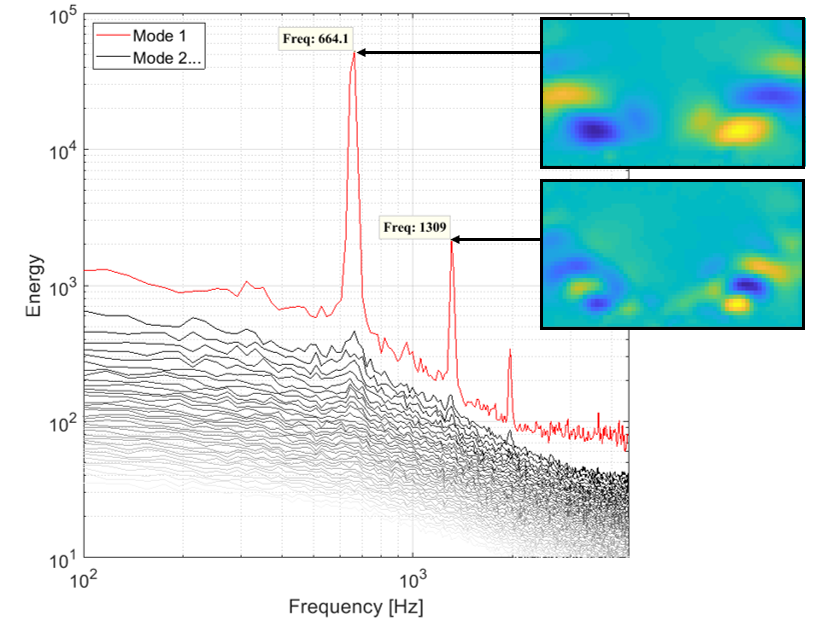}
    \caption{Radial mode shapes of the first mode for air split=0.2.}
    \label{fig:dlr_spod_0.2}
\end{figure}

Fig. \ref{fig:dlr_spod_0.2} shows the spatial mode shapes of the radial velocity at the dominant peak frequencies at the lowest air split (air split=0.2) case. The spatial structure of the first mode at $f=664.1$ Hz shows concentrated regions of fluctuation on either side of the central axis. Since this is a mode shape of radial velocity, this spatial structure corresponds to a symmetric oscillation, which is characteristic of the flow response to a longitudinal thermoacoustic oscillation; we will refer to this mode as an $m=0$ mode, where $m$ is the azimuthal wavenumber of the flow disturbance. In this case, the longitudinal acoustic oscillation provides a symmetric disturbance to the flow and so the flow responds symmetrically, as shown in previous swirling flow studies by \cite{oconnor2012transverse}. The mode at the second frequency, $f=1309$ Hz, is the first harmonic of that mode, which can be seen both by the doubling in frequency as well as the halving of the spatial wavelength of the oscillations in the mode shape. These mode shapes show that, in the lowest air split case, the main oscillation mode is the thermoacoustic oscillation. No PVC is seen to be present in this case, and this is supported by the weak recirculation zone seen in time-averaged velocity profiles in Fig. \ref{fig:dlr_uavg}. 

\par
\begin{figure}
    \centering
    \includegraphics[width=90mm]{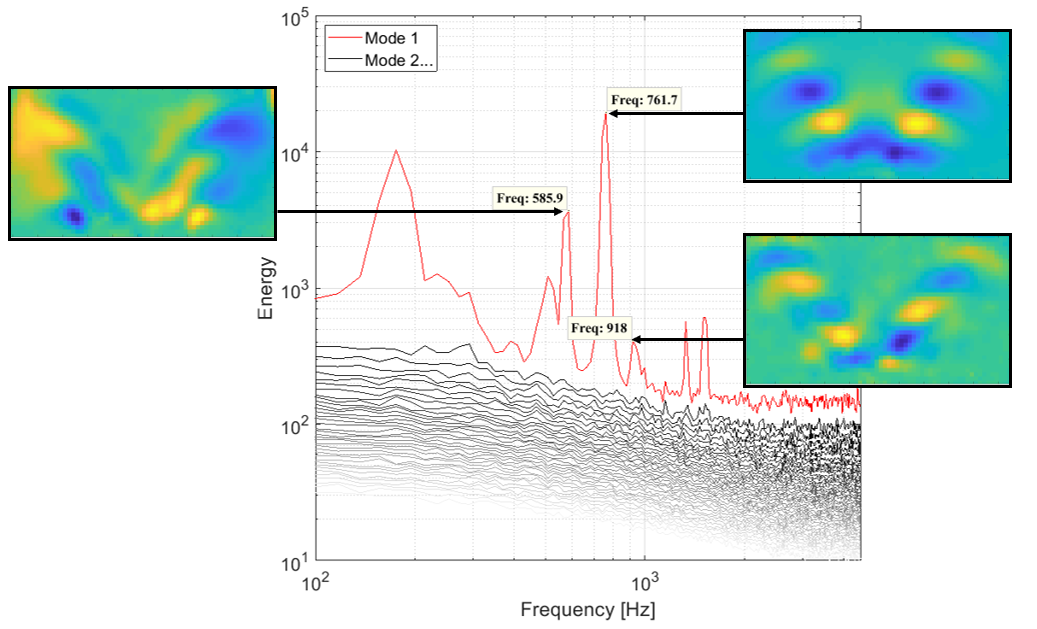}
    \caption{Radial mode shapes of the first mode for air split=0.5.}
    \label{fig:dlr_spod_0.5}
\end{figure}

As the air split increases, the strongest peak changes frequency and the mode shape of this new mode has a distinctively anti-symmetric spatial structure, indicative of a PVC mode shape \citep{manoharan2020PVC}. While the dominant peak now corresponds to the PVC mode, it is important to note that the thermoacoustic peaks are still present, but significantly weaker than the PVC. Figure \ref{fig:dlr_spod_0.5} shows, for the case with the highest air split of 0.5, the radial velocity mode shape of the strongest mode at a frequency of 762 Hz. This spatial mode shape shows a strong precessing center at the base and then helical vortical structures in the shear layer, both of which are characteristic of a $m=1$ PVC mode. The peak seen at 586 Hz still corresponds to a symmetric mode and so does the small peak seen at 918 Hz. Both these peaks always exhibit a symmetric spatial structure, characteristic of $m=0$ thermoacoustic modes. The spatial modes at the peak frequencies show that in the lower air split cases, the symmetric mode corresponding to thermoacoustic oscillation is dominant, but as the air split increases, the thermoacoustic modes are still present but the PVC gets stronger and dominates the flow field.

\par
\begin{figure}
    \centering
    \includegraphics[width=120mm]{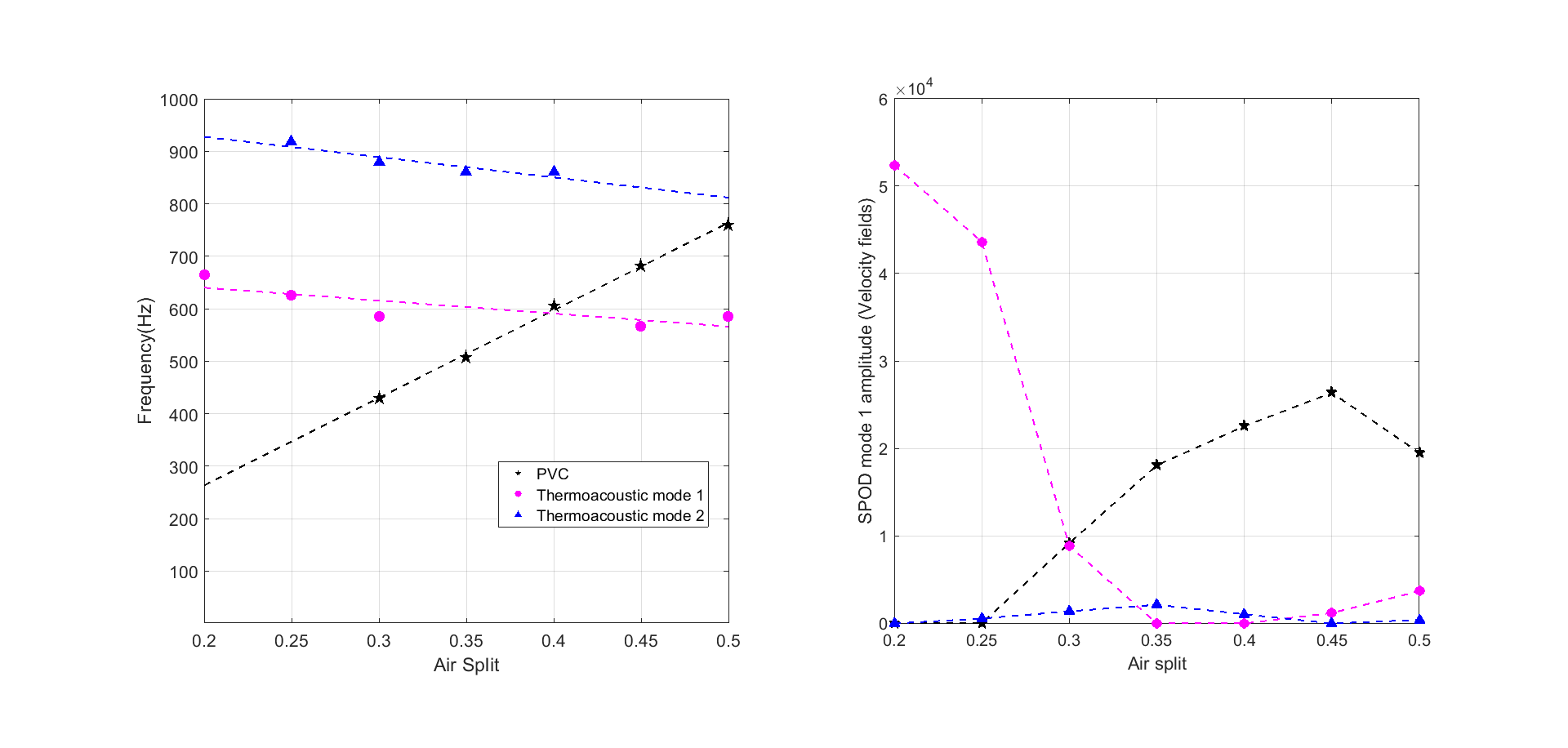}
    \caption{Frequency (left) and amplitude (right) characterization of the three oscillation modes seen at each air-split condition from SPOD.}
    \label{fig:dlr_freqs}
\end{figure}

The SPOD mode shapes are reconstructed at each of the frequency peaks in all the cases and the peaks are characterized by the spatial structure of the radial mode shape. From the mode shapes, it is seen that there are two symmetric $m=0$ modes in the flow corresponding to frequencies near 600 Hz and 800 Hz, both of these modes are thermoacoustic oscillation modes. In the lowest air split case, only the 600 Hz mode is seen, but subsequent cases show the second thermoacoustic mode as well, where the 600 Hz mode is always several orders of magnitude stronger than the 800 Hz thermoacoustic mode. The PVC $m=1$ mode strengthens with increasing air split and its frequency increases with linearly with increasing air split. The frequencies and amplitudes of three oscillation modes from the SPOD for each condition are depicted in the plot shown in Fig. \ref{fig:dlr_freqs}.

\par
From the plots in Fig. \ref{fig:dlr_freqs} and the eigenvalue spectra in Fig.\ref{fig:spod_velocity}, two important characteristics of the oscillation modes in the system can be observed. First, while the frequency of the PVC increases linearly with increasing air-split, the two symmetric frequencies remain largely the same at all air split conditions; this is further confirmation that the 600 and 800 Hz modes are thermoacoustic as they depend on the geometry of the system, not the flow conditions. Second, increasing amplitude of the PVC corresponds to decreasing amplitude of the first thermoacoustic mode; the second thermoacoustic mode is weak at all conditions. A key observation of the experimental portion of this study is that it appears that the global $m=1$ hydrodynamic instability, the precessing vortex core, suppresses the $m=0$ flow field response to thermoacoustic oscillation if the PVC amplitude is large enough. The theoretical formulation described next explains the reason for this observed behavior.

\par
\input{Theoretical_Formulation}

\section{Impact of hydrodynamics on flame behavior}
\label{sec:flame}
\begin{figure}
    \centering
    \includegraphics[width=90mm]{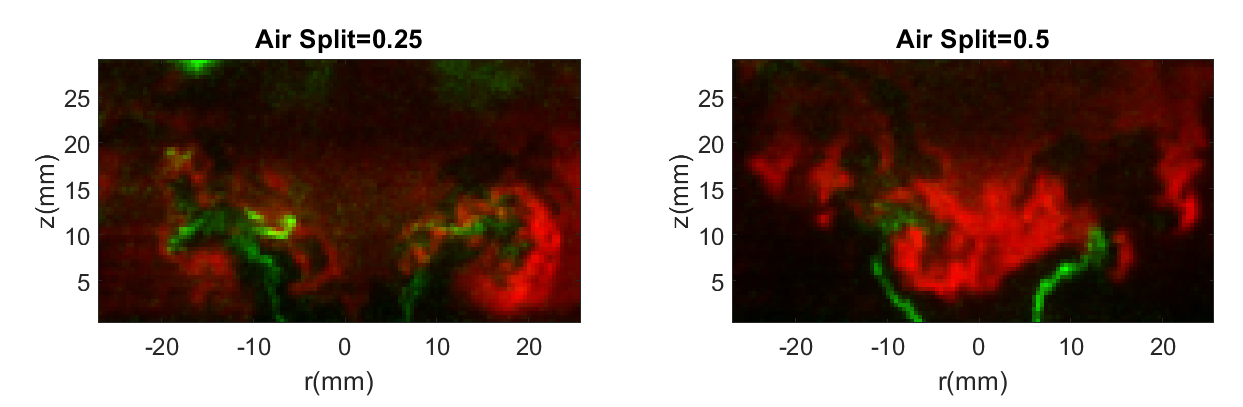}
    \caption{Instantaneous images of superimposed OH-PLIF (red) and acetone-PLIF (green) profiles at air splits 0.25 (left) and 0.5 (right). Videos of these conditions are provided in the supplemental material.}
    \label{fig:flame_fuel_snaphsot}
\end{figure}

The time-averaged fuel and flame contours, such as those shown in Fig. \ref{fig:dlr_oh_acetone_avg}, do not show significant variations with air split. The instantaneous images, however, distinctly show a shift from largely symmetric to asymmetric with increasing air split. Figure \ref{fig:flame_fuel_snaphsot} shows instantaneous images of the flame, obtained from the OH-PLIF images (depicted by the red contour), and the fuel distribution, obtained from the acetone-PLIF images (depicted by the green contour), at a low air split of 0.25 and a high air split of 0.5.

\par 
The instantaneous images illustrate the transition from symmetric motion to anti-symmetric motion as the air split is increased. In the low air split case, symmetric vortex roll-up of the flame surface can be seen. The fuel jets interact with the symmetric vortical structures, leading up to the onset of thermoacoustic instability as a result of periodic fluctuations in both the air and fuel flow rates. The high air split case, on the other hand, shows an anti-symmetric structure, which resembles the helical vortex structures generated by the PVC. These instantaneous images illustrate the influence that the PVC has on the flame structure and fuel distribution in the system. In the absence of the PVC, the symmetric thermoacoustic mode governs the flame response, but as the PVC amplitude increases, the PVC begins to govern the fuel flow and flame response. This transition would suggest that the flame response is strongly coupled to the velocity field in the flow, indicating that in the presence of a sufficiently strong global hydrodynamic instability, the flame is less receptive to thermoacoustic oscillations.
\par
The results from the frequency-domain characterization of the flow field show that as the air split is increased, the PVC strengthens and eventually dominates the flow field dynamics. The instantaneous flame structure and fuel distribution profiles indicate that the flame and fuel response are coupled with the velocity response. In order to draw conclusions about the impact of the PVC on the fuel/air mixing and flame dynamics, a more quantitative analysis is needed. 
\par
\begin{figure}
    \centering
    \includegraphics[width=100mm]{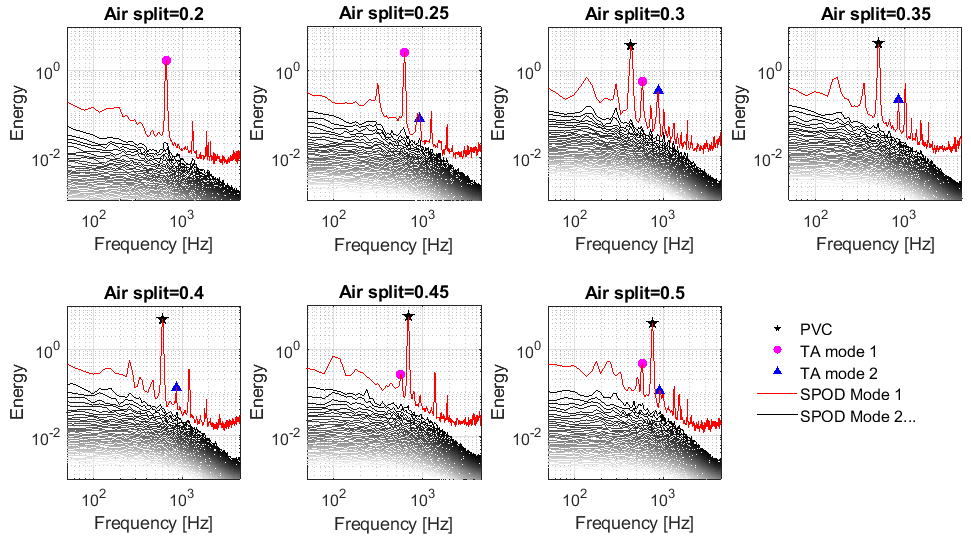}
    \caption{Eigenvalue spectra of OH-PLIF field obtained from spectral POD.}
    \label{fig:dlr_spod_oh}
\end{figure}

 
We perform spectral proper orthogonal decomposition on the acetone- and OH-PLIF image time series. Figure \ref{fig:dlr_spod_oh} shows the eigenvalue spectra of the flame oscillation, obtained from the OH-PLIF images, at each air split. The peak frequencies in the OH-PLIF spectra are similar to the peaks seen in the velocity spectra, where the thermoacoustic modes and PVC are indicated with the same symbols as in Figs. \ref{fig:dlr_acoustic} and \ref{fig:spod_velocity}. In the low air split cases, the thermoacoustic mode dominates the flame behavior as the first thermoacoustic mode amplitude is the highest peak. As the air split is increased, the PVC peak grows in strength, indicating that the PVC fluctuations drive the flame behavior. Furthermore, similar to the flow field response, whenever the PVC amplitude is high, the thermoacoustic mode is suppressed. The eigenvalue spectra of the OH-PLIF fields indicate that the flame is strongly coupled with the flow field.

\par
\begin{figure}
    \centering
    \includegraphics[width=100mm]{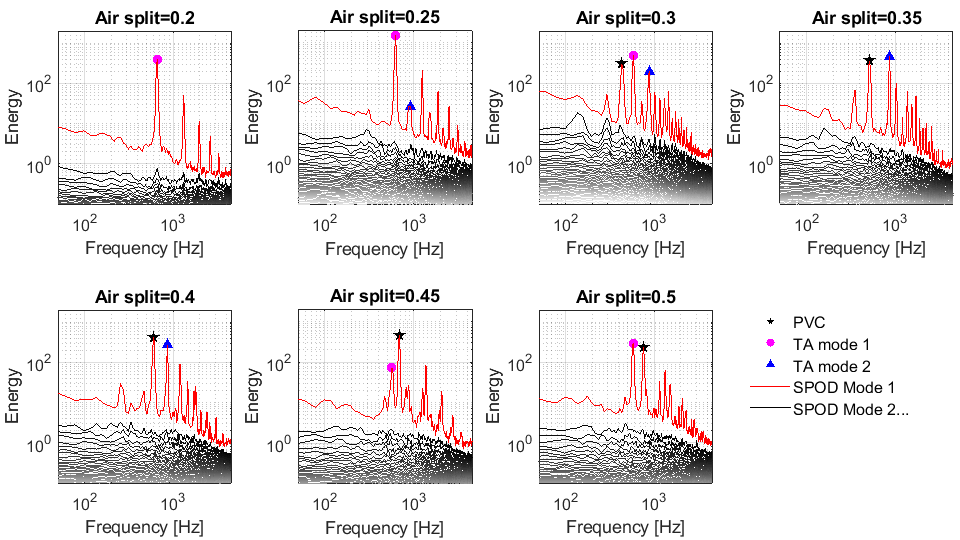}
    \caption{Eigenvalue spectra of the acetone field obtained from spectral POD.}
    \label{fig:dlr_spod_acetone}
\end{figure}

Figure \ref{fig:dlr_spod_acetone} shows the SPOD spectra of the acetone-PLIF images. Compared to the eigenvalue spectra of the velocity fields, the eigenvalue spectra of the PLIF data shows more peaks. This is likely because the acetone-PLIF data is inherently more noisy as compared to the velocity data. Unlike the OH-PLIF eigenvalue spectra, the acetone-PLIF eigenvalue spectra are distinctly different from the velocity field. While the peak frequencies are largely the same, the relative peak amplitudes are significantly different. Just like in the flame and velocity spectra, as the air split increases, the strength of the PVC mode increases. However, in the acetone spectra, the thermoacoustic mode is just as strong, if not stronger, than the PVC mode, even in the high air split cases, indicating that, unlike flame surface oscillation, fuel fluctuation is not dominated purely by the PVC. This is likely due to the distance between the base of the recirculation zone and the location of fuel injection at the dump plane, so the PVC does not impact fuel injection. As seen in Fig. \ref{fig:min_u}, the center of the recirculation zone is located at a downstream location of 10 mm, whereas the fuel is injected at the dump plane. The fuel fluctuation, therefore, is impacted more by the pressure oscillations, which likely have an anti-node at the dump plane, than the oscillation of the PVC.

\par
The eigenvalue spectra of the acetone-PLIF images suggest that the fuel fluctuation is not as sensitive to the PVC as the flame fluctuation. However, since the spectral POD method considers the energy of the entire flow field, it is important to be careful when interpreting the results of the eigenvalue decomposition. While the acetone burns out fairly close to the nozzle, the LIF signal further downstream could be signal from either flame or soot luminosity. In order to correctly interpret the results of the spectral characterization, the oscillation modes at different spatial locations in the flow need to be observed. To that end, we perform a Fourier transform on the time series of the PLIF images at every pixel location in the field of view. At each location, the peak frequency is extracted and plotted as a contour in space.   
\par
\begin{figure}
    \centering
    \includegraphics[width=120mm]{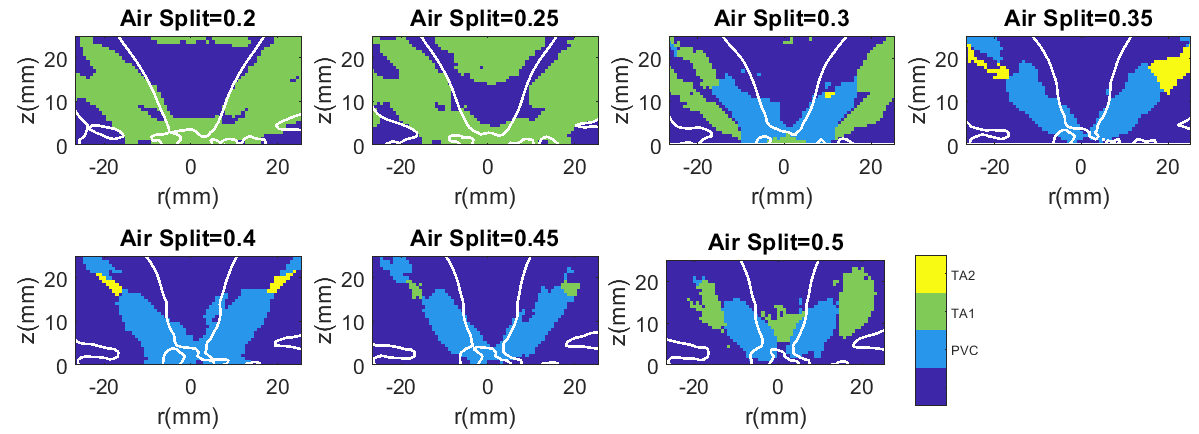}
    \caption{Peak frequency contours of OH field. The white contour represents the recirculation zone border, ($v_{z,mean}=0$).}
    \label{fig:dlr_contour_oh}
\end{figure}

Figure \ref{fig:dlr_contour_oh} shows the peak frequency contours of the OH field. The white contour represents the $v_{z,mean}=0$ contour, which is indicative of the edge recirculation zone; this inner shear layer is where the flame stabilizes in this configuration. The spatial peak frequency contours of the OH field, much like its SPOD spectra, show that the OH field is largely driven by the PVC once the PVC is of sufficient strength. The peak frequency contours of the velocity fields (not shown) are very similar to that of the OH field.

\begin{figure}
    \centering
    \includegraphics[width=120mm]{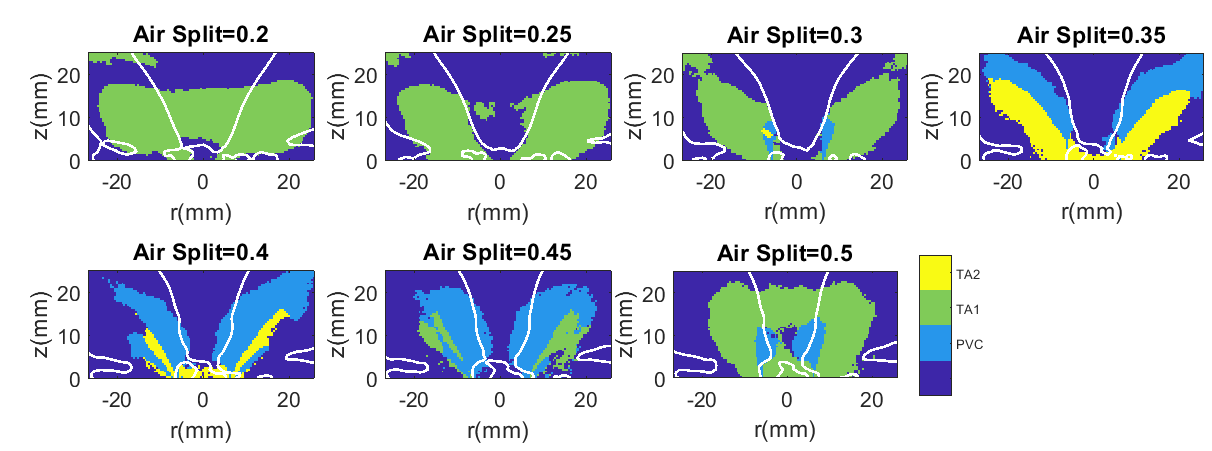}
    \caption{Peak frequency contours of acetone field. The white contour represents the recirculation zone border, ($v_{z,mean}=0$).}
    \label{fig:dlr_contour_acetone}
\end{figure}

\par
Figure \ref{fig:dlr_contour_acetone} shows the peak frequency contours of the acetone field. The acetone contours show a significantly different spatial distribution in the peak frequency contours compared to the frequency contours of the OH and velocity fields. While the PVC oscillation still dominates regions close to the shear layer, these regions are substantially smaller when compared to the peak frequency contours of the OH field. The results from this spatial frequency characterization support the results obtained from the eigenvalue spectra obtained using SPOD. As air split increases, the PVC strengthens and progressively dominates the flame and flow response. However, it is seen that, due to the relative offset between the downstream locations of the recirculation zone and the fuel injection location, the fuel fluctuation is less sensitive to the PVC oscillation when compared to the flame and flow fields. 
\par 
These differences in dominant frequency have implications for thermoacoustic coupling processes. In this complex, realistic combustor flow field, operating at an elevated pressure, the thermoacoustic feedback loop could involve both velocity coupling and equivalence ratio coupling mechanisms. As such, it is important to characterize not only how these coupling mechanisms evolve with changing flow conditions but also how the two coupling pathways interact with one another. The results presented here indicate that the hydrodynamic instability and equivalence ratio coupling pathways do not strongly interact with each other since the recirculation zone is located sufficiently further downstream of the fuel injection location.

\section{Conclusions}
This study showed using both experimental and theoretical analysis that a precessing vortex core can suppress longitudinal thermoacoustic instability in a partially-premixed, swirl-stabilized flame at elevated pressure. When the PVC frequency, which increases with air split, approaches that of the thermoacoustic instability, a nonlinear coupling, shown by eq. \ref{eq:A0_evol}, between the $m=1$ PVC motion and the $m=0$ shear layer response leads to suppression of the $m=0$ mode and hence the velocity-coupled thermoacoustic instability. 
\par
In addition to this main conclusion, this work has identified a number of new results in the PVC literature. First, varying the air split in this injector does not change the swirl number of the flow as the geometric swirl number of the inner and outer swirl passages are relatively similar. However, increases in air split result in stronger central recirculation in the vortex breakdown bubble, although the location of the bubble does not change significantly with air split. This result is explained by the variation in the characteristic critical swirl number of the burner that results from changes in air split. This critical swirl number is the minimum swirl number needed at a given air split to initiate vortex breakdown in the flow. The results suggest that increasing the air split decreases the critical swirl number as a result of structural changes to the flow field. As the offset between the operating swirl number of the experiment and critical swirl number increases, the vortex breakdown bubble recirculation strengthens. This result has been seen in previous studies by increasing swirl number \citep{liang_maxworthy05,escudier1985}, but, to the best of the authors' knowledge, not seen by decreasing the critical swirl number at constant swirl.
\par
Second, the impact of the PVC on shear layer behavior is consistent with many previous studies \citep{oberleithner2011stabilityanalysis,oconnor18PVCshear,moeck2012PVCandAcoustics}, where the shear layers surrounding the vortex breakdown region respond to the wavemaker region at the base of the recirculation zone driven by the $m=1$ precession mode of the breakdown bubble. The spatial plots of peak frequency in Fig. \ref{fig:dlr_contour_acetone} show, however, that the influence of the PVC is localized and does not drive fluctuations in the fuel at the dump plane, which is located at a distance from the upstream stagnation point of the recirculation zone. This result indicates that the influence of the PVC is local, unlike the dynamics of the thermoacoustic mode, which are felt throughout the combustor as they are driven by a pressure mode shape inside the combustor.
\par
Finally, the insights obtained from this analysis have significant implications for the suppression of thermoacoustic oscillations in gas turbine combustors. Work by first \cite{mathews2016PVC} and then \cite{oconnor18PVCshear} showed, in a non-reacting, variable-swirl-number jet configuration, that the presence of the PVC suppressed the response of the shear layer to external forcing. Calculation of the spatial growth rate of the $m=0$ shear layer instability in \cite{oconnor18PVCshear} indicated that the PVC caused the growth rate to become negative, resulting in no response at the forcing frequency. We had suggested, at the time, that these results may be extended to reacting circumstances as long as the PVC could be sustained. The present results confirm this hypothesis by showing the nonlinear coupling mechanism that leads to $m=0$ mode suppression, and hence suppression of the velocity-coupling pathway for thermoacoustic instability. This result suggests that ``designing" a PVC in a combustor flow field at frequencies close to those of a longitudinal thermoacoustic mode could suppress oscillations.

\section{Acknowledgements}
This work was supported by National Science Foundation grant CBET-1749679 and Air Force Office of Scientific Research grant FA9550-16-1-0044. Any opinions, findings, and conclusions or recommendations expressed in this material are those of the authors and do not necessarily reflect the views of the National Science Foundation or the Air Force Office of Scientific Research. The assistance of Dr. Jisu Yoon (ADD, Republic of Korea), Dr. Carrie Noren (AFRL, USA) and Dr. Klaus-Peter Geigle (DLR, Germany) with the setup and operation of the experimental apparatus described in this paper are gratefully acknowledged.

\section{Declaration of Interests}
The authors report no conflict of interest.

\appendix

\section{Governing equations in the operator form}
\input{Appendix_A}

\section{Coefficients in the evolution equations of $A_{1}$ and $A_{0}$}
\input{Appendix_B}

\section{Operator definitions}
\input{Appendix_C}

\bibliographystyle{agsm}
\bibliography{References_JFM2020}

\end{document}

%% file: Theoretical_Formulation.tex
\section{Theoretical Formulation}
\label{sec:theory}
We formulate an asymptotic solution to the variable-density, low Mach number Navier-Stokes equations to gain insight into the underlying mechanisms that determine the coherent flow oscillation behavior observed in the experimental results. The SPOD analysis presented in the previous section suggests that acoustic forcing from the first thermoacoustic mode excites an axisymmetric \textit{hydrodynamic} response at a fixed frequency. The response is hydrodynamic because it is driven by the dynamics of the vorticity field when perturbed by acoustic velocity oscillations. Therefore, it is reasonable to expect that these hydrodynamic oscillations are strong in regions of high time-averaged vorticity. In regions where the time-averaged vorticity is small, flow oscillations are simply acoustic velocity oscillations generated by the acoustic pressure field. The discussion in prior sections has shown that the amplitude of axisymmetric flow oscillations are strong in the flame region and smaller in regions away from the flame. This shows that the thermoacoustic pressure oscillation imposes a weak forcing on the time-averaged flow, where `weak' implies that the amplitude of the \textit{hydrodynamic} response is much larger than that of the imposed forcing. This is common in acoustically-forced shear flows \citep{mathews2016PVC}.
\par
\cite{manoharan2020PVC} showed, using a high-$Re$, variable-$S$ jet, that PVC oscillations are a result of linear instability at a critical swirl number $S_{c}$ that induces a stable limit cycle oscillation in the flow, characterized by helical precession of the vortex breakdown bubble around the flow central axis. This precession, in turn, results in the helical perturbation and rollup of the shear layers. A large number of prior experiments support this observation \citep{oberleithner2011stabilityanalysis,anacleto2003,escudier1985}. The value of $S_{c}$ depends on the evolution of the time-averaged flow field as $S$ varies. Therefore, in the case of the present experiment, varying the air split between the inner and outer flow circuits has the impact of varying $S_{c}$ while keeping $S$ nominally constant, as can be seen in Fig. \ref{fig:min_u} (right). Accordingly, for an air split of 0.3 and above, Fig.~\ref{fig:spod_velocity} shows that $S>S_{c}$ due to the emergence of the PVC. 
\par 
Following \cite{manoharan2020PVC}, we derive an asymptotic solution for the flow state at a given $S$ assuming a small departure from $S_{c}$ as follows,
\begin{equation}
\label{eq:S_Sc_rel}
    S = S_c + \epsilon^2 \Delta_s 
\end{equation}
where $\epsilon$ is a small number and $\Delta_s \sim O(1)$. The governing equations are the nonlinear, variable density Navier-Stokes (NS) equations in the low Mach number limit, formulated in a cylindrical coordinate $(r,\theta,z)$ system with the $z$ and $r$ axes aligned like the $x$ and $y$ axes, respectively, as shown in Fig.~\ref{fig:dlr_experiment}. Axial $(u_z)$ and radial $(u_r)$ velocity components are non-dimensionalized using a suitably chosen reference velocity, $U_{z,ref}$. The azimuthal $(u_\theta)$ velocity component is non-dimensionalized using $U_{\theta,ref}=SU_{z,ref}$, where $S$ is the flow swirl number. All lengths are normalized by the burner outer diameter. The governing equations for the coherent flow component, $\pmb{\tilde{q}} (r,\theta,z,t)= \big[ \tilde{\rho}, \tilde{u}_r, \tilde{u}_\theta, \tilde{u}_z, \tilde{p} \big]^T$, can now be written as follows,
\begin{equation}
\label{eq:Evol_equ_q_tilde}
\begin{split}
	\Bigg(
	\mathscr{B}
	+
	\mathscr{B}^S\{\pmb{\tilde{q}}\}
	\Bigg)
	\frac{\partial \pmb{\tilde{q}}}{\partial t}
	=
	-
	\mathscr{N}\{\pmb{\tilde{q}}\} \pmb{\tilde{q}}
	-
	S \mathscr{N}^S\{\pmb{\tilde{q}}\} \pmb{\tilde{q}}
	-
	S^2 \mathscr{N}^{SS}\{\pmb{\tilde{q}}\} \pmb{\tilde{q}}
	\\
	+
	\mathcal{L}_T \pmb{\tilde{q}}
	+
	S \mathcal{L}_T^S \pmb{\tilde{q}}
	+
	\epsilon^3 \pmb{\hat{q}_a} (r,z)
    \cos{(\omega_a t)}
\end{split}
\end{equation}
where, the $O(\epsilon^3)$ term on the right is the forcing term capturing the weak forcing imposed by the axisymmetric thermoacoustic pressure oscillation with a frequency $\omega_a$ and spatial amplitude distribution $\pmb{\hat{q}_a}$.
\par
In eq. \ref{eq:Evol_equ_q_tilde}, the operators $\mathscr{B}$ and $\mathscr{B}^S \{ \pmb{\tilde{q}} \}$ are diagonal matrices with elements $\{1,0,0,0,0\}$ and $\{0, \rho, S \rho, \rho,0 \}$ respectively and $\mathscr{N} \{ \pmb{\tilde{q}} \}$, $\mathscr{N}^S \{ \pmb{\tilde{q}} \}$ and $\mathscr{N}^{SS} \{ \pmb{\tilde{q}} \}$ are nonlinear operators arising from convective terms. $\mathcal{L}_T$ and $\mathcal{L}^S_T$ in eq. \ref{eq:Evol_equ_q_tilde} are the linear operators arising from the pressure gradient term and molecular and turbulent transport terms. An eddy viscosity model along with the Boussinesq assumption is used to close the turbulent transport term as in prior studies \citep{manoharan2020PVC,juniper_2016,oberandstohr15pvc}. The various operators in eq.~\ref{eq:Evol_equ_q_tilde} are functions of the quantities presented within the associated  braces, $\{ ~\}$. Also, solutions to eq.~\ref{eq:Evol_equ_q_tilde} must be periodic in $\theta$ and obey no-slip conditions at the walls and kinematic compatibility conditions on the flow centerline \citep{batchelor_gill_1962}. The details of the all the operators used in eq. \ref{eq:Evol_equ_q_tilde} are algebraically complex expressions and are therefore presented in the appendix (eqs. \ref{App_eq:N_oper_dfn}, \ref{App_eq:NS_oper_dfn} and \ref{App_eq:NSS_oper_dfn}).
\par
We note from the discussion in prior sections that the flow state is characterized by flow oscillations with constant amplitudes. Accordingly, we now derive a solution to eq.~\ref{eq:Evol_equ_q_tilde} accurate up to $O(\epsilon^3)$, composed of oscillatory components whose amplitudes grow slowly to their steady-state oscillation values, using the method of multiple scales \citep{nayfeh2008perturbation}. Thus, we introduce a `fast' time scale $t_1 = t$, a `slow' time scale $t_2 = \epsilon^2 t$, and an expansion for $\pmb{\tilde{q}}$ in terms of $\epsilon$ as follows,
\begin{equation}
\label{eq:q_asymp_expans}
    \pmb{\tilde{q}} (r,\theta,z,t_1,t_2)
    =
    \pmb{q_0} (r,z)
	+
	\epsilon \pmb{q_{1}} (r,\theta,z,t_1,t_2)
	+
	\epsilon^2 \pmb{q_{2}} (r,\theta,z,t_1,t_2)
	+
	\epsilon^3 \pmb{q_{3}} (r,\theta,z,t_1,t_2)
	+
	...
\end{equation}
where, $\pmb{q_0}$ is the time-averaged flow state at $S_{c}$ and each term in the expansion is assumed to depend independently on fast and slow time scales. All the operators in eq. \ref{eq:Evol_equ_q_tilde} are also expanded in powers of $\epsilon$ (see eqs. \ref{App_eq:N_oper_expans}, \ref{App_eq:NS_oper_expans} and \ref{App_eq:NSS_oper_expans} for details). We then substitute these expansions into eq.~\ref{eq:Evol_equ_q_tilde} and compare terms that are coefficients of the same power of $\epsilon$ on both sides. This yields a sequence of equations for $\pmb{q_{1}}$, $\pmb{q_{2}}$, $\dots$, etc. Although not shown here, boundary conditions are treated similarly and yield homogeneous boundary conditions that must be obeyed by the coefficients in eq.~\ref{eq:q_asymp_expans} at each order of $\epsilon$.
\par 
At $O(\epsilon)$, the unsteady, linearized Navier-Stokes (LNS) equations for $q_1$ are obtained as follows,
\begin{equation}
\label{eq:Order_eps}
    \left(
    \mathscr{B}_0 \frac{\partial }{\partial t_1}
    +
    \mathscr{L}
    \right)
    \pmb{q_1}
    =
    0
\end{equation}
where, $\mathscr{L}$ is an operator representing the spatial derivative terms in the LNS equations and depends on $\pmb{q_{0}}$ and $S_{c}$. The details of the operators $\mathscr{B}_0$ and $\mathscr{L}$ are presented in the appendix (eqs. \ref{App_eq:B0_oper_dfn} and \ref{App_eq:L_oper_dfn} respectively). We decompose $\pmb{q_{1}}$ into Fourier modes in the azimuthal direction and write the solution to eq. \ref{eq:Order_eps} as follows, 
\begin{equation}
\label{eq:Order_eps_sol}
    \pmb{q_1} (r,\theta,z,t_1,t_2)
    =
    A_1(t_2)
    \pmb{\hat{q}_{1}}(r,z)
    e^{i (\theta - \omega_{1} t_1)}
    +
    A_0(t_2)
    \pmb{\hat{q}_{0}}(r,z)
    e^{-i \omega_{0} t_1}
    +
    c.c.
\end{equation}
where, for the present flow, contributions from an axisymmetric mode ($\pmb{\hat{q}_{0}}$) and a helical mode ($\pmb{\hat{q}_{1}}$) have been retained. Thus, $\omega_{0}$ and $\omega_{1}$ are their associated natural frequencies, which are given by the solution to the following eigenvalue problem, 
\begin{equation}
\label{eq:Sol_homo_equ}
    \left(
    - i \omega_{m} \mathscr{B}_0 
    +
    \mathscr{L}_m
    \right)
    \pmb{\hat{q}_{m}} (r,z)
    =
    0
\end{equation}
where, the operator $\mathscr{L}_m$ is obtained from $\mathscr{L}$ by the transformation $\partial/\partial\theta\rightarrow im$ and $m=0,1$ for the axisymmetric and helical contributions in eq.~\ref{eq:Order_eps_sol}, respectively. The fact that the axisymmetric hydrodynamic mode oscillation frequency does not depend on air split, as the data analysis in prior sections shows, reveals that this mode is not self-excited, i.e., $\omega_{0i}<0$ - the subscript `$i$' represents the imaginary part. Accordingly, the thermoacoustic frequency, $\omega_{a}$ can now be written in terms of $\omega_{0}$ and the difference between $S$ and $S_{c}$ as follows,
\begin{equation}
\label{eq:forcing_freq_omega0_reln}
    \omega_a
    =
    \omega_0
    +
    \epsilon^2 \Omega_a
\end{equation}
where, $\Omega_{a}\sim O(1)$ and quantifies the extent of de-tuning between the natural oscillation frequency $\omega _{0}$ and the thermoacoustic frequency $\omega_{a}$. Next, since $S_{c}$ is the swirl number at which PVC oscillations originate, $\omega_{1}$ is neutrally stable, i.e. $\omega_{1i}=0$~\citep{manoharan2020PVC}. 
\par
The solution for $\pmb{q_{2}}$ in eq.~\ref{eq:q_asymp_expans} is determined from the equation obtained by comparing $O(\epsilon^2)$ terms on both sides of eq.~\ref{eq:Evol_equ_q_tilde} as follows,
\begin{equation}
\label{eq:Order_eps2_gen_sol}
\begin{split}
    \pmb{q_2} (r,\theta,z,t_1,t_2)
    =
    \Delta_s \pmb{\hat{q}_\Delta}
    +
    |A_1|^2 \pmb{\hat{q}_{A_1 A_1^{*}}}
    ~~~~~~~~~~~~~~~~~~~~~~~~~~~~~~~~~~~~~~~~~~~~~~~~~~~~~~~~~~~~~~~~~~~~~~~~~~~~~~~~~~
    \\
    ~~~~~~~~~~~~~~~~~~~
    +
    \Big(
    A_1 A_0 \pmb{\hat{q}_{A_1 A_0}} e^{i \theta} e^{-i(\omega_{1}+\omega_{0}) t_1} 
    +
    A_1 A^{*}_0 \pmb{\hat{q}_{A_1 A_0^{*}}} e^{i \theta} e^{-i(\omega_{1}-\omega^*_{0}) t_1}
    +
    c.c.
    \Big)
    ~~~~~~~~~~~~~~~~~~~~~
    \\
    +
    \Big(
    (A_1)^2 \pmb{\hat{q}_{A_1 A_1}} e^{2i(\theta - \omega_{1} t_1)}
    +
    (A_0)^2 \pmb{\hat{q}_{A_0 A_0}} e^{-2i \omega_{0} t_1}
    +
    c.c.
    \Big)
    ~~~~~~~~~~~~~~~~~~~~~~~~~~~~~~~~~~~~
    \\
    +
    \Big(
    B_1
    \pmb{\hat{q}_{1}}
    e^{i (\theta - \omega_{1} t_1)}
    +
    B_0
    \pmb{\hat{q}_{0}}
    e^{-i \omega_{0} t_1}
    +
    c.c.
    \Big)
    ~~~~~~~~~~~~~~~~~~~~~~~~~~~~~~~~~~~~~~~~~~~~~~~~~~~~~~~
\end{split}
\end{equation}
where, the first term quantifies the change in the time-averaged flow from its state at $S_{c}$, the second term is the time-averaged distortion (to leading order) due to finite amplitude helical oscillations. The third set of terms are due to nonlinear coupling between the helical and axisymmetric modes. Note that this results in helical oscillations at frequencies corresponding to the sum and difference of the thermoacoustic and PVC frequencies. The fourth set of terms are the first harmonic of the helical and axisymmetric oscillations. The last set of terms are the complementary functions of the homogeneous form of eq.~\ref{eq:Evol_equ_q_tilde} at $O(\epsilon^2)$. These terms are included here for mathematical completeness and will not contribute to the final steady state oscillatory solution up to $O(\epsilon^3)$. 


\par
The amplitudes $A_{0} (t_2)$ and $A_1 (t_2)$ are determined from the equation for $\pmb{q_{3}}$ (coefficient of $\epsilon^3$ in eq.~\ref{eq:q_asymp_expans}). This equation has oscillatory source terms of the form $e^{\-i\omega_{0} t_{1}}$ and $e^{-i\omega_{1} t_{1}}$ which in general yield terms in the solution for $q_{3}$ whose value grows exponentially with time. Thus, a bounded solution is possible only when the coefficients of these secular terms vanish. This condition yields evolution equations for $A_0(t_2)$ and $A_1(t_2)$ as follows, 
\begin{equation}
\label{eq:A0_evol}
    \frac{d A_0}{d t_2}
    =
    \Delta_s \alpha_{A_0} A_0
    -
    \beta_{A_0 A_1} |A_1|^2
    A_0
    +
    \beta_{A_0 f} e^{-i \Omega_a t_2}
\end{equation}
\begin{equation}
\label{eq:A1_evol}
    \frac{d A_1}{d t_2}
    =
    \Delta_s \alpha_{A_1} A_1
    -
    \beta_{A_1 A_1} |A_1|^2
    A_1
\end{equation}
where, the coefficients in eqs.~\ref{eq:A0_evol} and \ref{eq:A1_evol} are determined from inner products between the \textit{adjoint} modes $\pmb{\hat{q}_0^{\dagger}}$ and $\pmb{\hat{q}_1^{\dagger}}$, and expressions involving various the functions appearing in eqs.~\ref{eq:Order_eps_sol} and \ref{eq:Order_eps2_gen_sol}. These expressions are given in the appendix (eqs. \ref{App_eq:alpha_A1} - \ref{App_eq:Beta_A0_A1}). Equation~\ref{eq:A1_evol} is equivalent to that derived by \cite{manoharan2020PVC} for the evolution of helical mode oscillations in an unforced flow with constant density. This shows that in the limit of weak axisymmetric forcing, the evolution of $A_1$, i.e., helical oscillations, is independent of $A_0$, i.e., the hydrodynamic response to the thermoacoustic forcing. On the other hand, the second term in eq.~\ref{eq:A0_evol} shows that helical oscillations can influence the evolution of the axisymmetric hydrodynamic response. The efficiency with which this occurs is determined by the value of $\beta_{A_0 A_1}$.
\par 
The steady-state oscillation amplitudes are determined as follows. First, we introduce  $A_0(t_2) = D_{0}(t_2) \exp[i\phi_{0}(t_{2})]$ and $A_1(t_{2}) = D_{1}(t_{2}) \exp[i\phi_{1}(t_{2})]$ in eqs.~\ref{eq:A0_evol} and \ref{eq:A1_evol}, and equate real and imaginary parts on both sides. The real parts define evolution equations for $D_{1}$ and $D_{0}$. Requiring time derivatives of $D_{0}$ and $D_{1}$ to vanish in these equations yields solutions for the steady-state amplitudes. The imaginary parts then yields expressions for the natural frequencies of the axisymmetric and helical modes at $S$ to leading order in $S-S_{c}$. 
\par 
The solutions for the steady-state amplitude of the helical mode and its characteristic frequency are as follows,
\begin{equation}
\label{eq:A_PVC_Sol}
    A_{PVC}
    =
    \sqrt{
    \frac{(S-S_c) \alpha_{A_1 r}}
    {\beta_{A_1 A_1 r}}
    }
\end{equation}
and,
\begin{equation}
\label{eq:f_PVC}
    \omega_{PVC}
    =
    \omega_{1}
    +
    (S-S_{c})
    \bigg(
    \alpha_{A_1 r}
    \frac{\beta_{A_1 A_1 i} }
    {\beta_{A_1 A_1 r}}
    -
    \alpha_{A_1 i}
    \bigg)
\end{equation}
where, the letters `$r$' and `$i$' in the subscripts denote real and imaginary parts. Note that as may be expected, the results in eqs.~\ref{eq:A_PVC_Sol} and \ref{eq:f_PVC} are analogous to those derived by \cite{manoharan2020PVC} for a constant density flow. 
\par 
\cite{manoharan2020PVC} show in the case of their variable-swirl, non-reacting jet that helical oscillations occur due to the inception of vortex breakdown at $S=S_{c}$, with the breakdown bubble being increasingly well-defined thereafter. For the present case, $S_{c}$ quantifies the minimum swirl intensity that the nozzle needs to generate for a given air split in order to induce vortex breakdown in the flow. Thus, the appearance of a well-defined vortex breakdown bubble on the flow centreline with increasing air split as Fig.~\ref{fig:min_u} (left) shows, with $S$ remaining nearly the same across air splits (see Fig.~\ref{fig:min_u} (right)), is a consequence of a corresponding reduction in $S_{c}$.  Therefore, eq.~\ref{eq:f_PVC} suggests the linear rise in characteristic PVC frequency with air split, as Fig.~\ref{fig:dlr_freqs} (left) shows, is due to the increase in $S-S_{c}$. Likewise, eq.~\ref{eq:A_PVC_Sol} then shows that the increase in PVC amplitude seen in Fig.~\ref{fig:dlr_freqs} (right) is because of the same reason. The slight decrease in the amplitude for an air split of 0.5 in Fig.~\ref{fig:dlr_freqs} (right) is possibly due to the fact that the bubble moves upstream into the nozzle, resulting in a part of the oscillation field moving out of the field of view of the sPIV cameras. This blocked view causes only a part of the total energy of the oscillation to be captured by the SPOD for this case alone. Note that the characteristic PVC frequency continues to increase linearly even for this case; see Fig.~\ref{fig:dlr_freqs} (left).
\par 
For the axisymmetric mode, the steady-state oscillation frequency is simply the thermoacoustic frequency $\omega_{a}$. The steady state amplitude, $A_{TH}$ of the axisymmetric mode is given by the following,
\begin{equation}
\label{eq:A_TH_Sol}
    A_{TH}
    =
    \frac{|\beta_{A_0 f}|}
    {\Big| i \big[ \omega_a - \omega_0 - i(S-S_{c}) \alpha_{A_{0}} \big] - \beta_{A_0 A_1} A_{PVC}^2 \Big|
    }
\end{equation}
where $\beta_{A_0 f}$ is given as follows:
\begin{equation}
\label{eq:Beta_A0_f}
    \beta_{A_0 f} 
    = 
    \frac
    {\langle \pmb{\hat{q}_0^{\dagger}}, \pmb{\hat{q}_a} \rangle}
    {2 \langle \pmb{\hat{q}_0^{\dagger}}, \mathscr{B}_0 \pmb{\hat{q}_{0}} \rangle}
\end{equation}
and $\beta_{A_0 A_1}$ is given by the following expression:
\begin{equation}
\begin{split}
\label{eq:Beta_A0_A1}
    \beta_{A_0 A_1}
    =
    \frac{1}
    {
    \langle 
    \pmb{\hat{q}_0^{\dagger}}
    ,
    \mathscr{B}_0
    \pmb{\hat{q}_{0}}
    \rangle
    }
    \bigg\langle
    \pmb{\hat{q}^{0\dagger}}
    ,
    \Big[
    \mathbb{I}_{(-1,1)} \{ \pmb{q_0}, \pmb{\hat{q}_1^{*}} \}
    \pmb{\hat{q}_{A_1 A_0}}
    +
    \mathbb{I}_{(1,-1)} \{ \pmb{q_0}, \pmb{\hat{q}_{A_1 A_0}} \}
    \pmb{\hat{q}_1^{*}}
    +
    \mathbb{I}_{(-1,1)} \{ \pmb{q_0}, \pmb{\hat{q}_{A_1^{*} A_0}} \}
    \pmb{\hat{q}_{1}}
    \\
    +
    ~
    \mathbb{I}_{(1,-1)} \{ \pmb{q_0}, \pmb{\hat{q}_{1}} \}
    \pmb{\hat{q}_{A_1^{*} A_0}}
    +
    \mathbb{I}_{(0,0)} \{ \pmb{q_0}, \pmb{\hat{q}_{A_1 A_1^{*}}} \}
    \pmb{\hat{q}_{0}}
    +
    \mathbb{S}_{(0,0)} \{ \pmb{q_0}, \pmb{\hat{q}_{0}} \}
    \pmb{\hat{q}_{A_1 A_1^{*}}}
    ~~~~~~~~~
    \\
    +
    ~
    \mathbb{Q}_{0} \{ \pmb{q_0}, \pmb{\hat{q}_1^{*}}, \pmb{\hat{q}_{1}} \}
    \pmb{\hat{q}_{0}}
    +
    \mathbb{Q}_{1} \{ \pmb{q_0}, \pmb{\hat{q}_1^{*}}, \pmb{\hat{q}_{0}} \}
    \pmb{\hat{q}_{1}}
    +
    \mathbb{Q}_{-1} \{ \pmb{q_0}, \pmb{\hat{q}_{1}}, \pmb{\hat{q}_{0}} \}
    \pmb{\hat{q}_1^{*}}
    ~~~~~~~~~~~~~~~~~~~~
    \\
    +
    \mathbb{T}_0
    \{
    \pmb{q_{0}}, 
    \pmb{\hat{q}_{0}},
    \pmb{\hat{q}_{1}},
    \pmb{\hat{q}_1^*},
    \pmb{\hat{q}_{A_1 A_1^*}},
    \pmb{\hat{q}_{A_1 A_0}},
    \pmb{\hat{q}_{A_1^* A_0}}
    \}
    \pmb{q_{0}}
    \Big]
    \bigg\rangle
    ~~~~~~~~~~~~~~~~~~~~~~~~~~~~~~~~~~~~~
\end{split}
\end{equation}
\\
where, the inner product between two functions $\pmb{\hat{p}_1} (r,z)$ and $\pmb{\hat{p}_2} (r,z)$ is defined as follows,
\begin{equation}
\label{eq:InnerProd_dfn}
    \langle
    \pmb{\hat{p}_1}, \pmb{\hat{p}_2}
    \rangle
    =
    \int_0^Z
    \int_0^R
    (\pmb{\hat{p}_1})^H
    \pmb{\hat{p}_2}
    \, r \, dr \, dz
\end{equation}
where, the superscript \lq{$H$}\rq~ denotes the transpose conjugate.
\par
In eq. \ref{eq:A_TH_Sol}, $\beta_{A_0 f}$ quantifies the receptivity of the axisymmetric mode to the forcing imposed by the thermoacosutic oscillations. The denominator in eq. \ref{eq:A_TH_Sol} quantifies two separate sources of de-tuning. The first term in box brackets in the denominator of eq. \ref{eq:A_TH_Sol} is the difference between the thermoacoustic forcing and the natural frequency of the axisymmetric mode at $S$. Thus, as may be expected, a larger difference between these two frequencies results in a smaller response amplitude. The last term in denominator of eq.~\ref{eq:A_TH_Sol} shows that the presence of helical oscillations can additionally reduce $A_{TH}$ if $|\beta_{A_{0} A_{1}}|$ is large, i.e., the nonlinear coupling between helical and axisymmetric oscillations is large. From the expression for $\beta_{A_0 A_1}$ in eq. \ref{eq:Beta_A0_A1}, the contribution from nonlinear coupling between the helical and axisymmetric mode depends on $\pmb{\hat{q}_{A_1^* A_0}}$ and $\pmb{\hat{q}_{A_1 A_0}}$. These quantities are components of the solution for $\pmb{q_{2}}$ (see eq.~\ref{eq:Order_eps2_gen_sol}) that oscillate with frequencies corresponding to the difference and sum of the helical and axisymmetric oscillations. The equation that determines $\pmb{\hat{q}_{A_1^* A_0}}$ is as follows, 
\begin{equation}
\label{eq:q_A1*_A0}
\begin{split}
    \Big(
    i \Delta\omega
    \mathscr{B}_0
    +
    \mathscr{L}_{-1}
    \Big)
    \pmb{\hat{q}_{A_1^{*} A_0}}
    =
    -
    \Big(
    \mathbb{I}_{(0,-1)} \{ \pmb{q_0}, \pmb{\hat{q}_0} \}
    \pmb{\hat{q}_1^{*}}
    +
    \mathbb{I}_{(-1,0)} \{ \pmb{q_0}, \pmb{\hat{q}_1^{*}} \}
    \pmb{\hat{q}_0}
    +
    \mathbb{Q}_0 \{ \pmb{q_0}, \pmb{\hat{q}_0}, \pmb{\hat{q}_1^{*}} \} \pmb{q_0}
    \Big)
\end{split}
\end{equation}
where, $\Delta\omega=\omega_{1}-\omega_{0}$ and the various operators on the right are given in the appendix (eqs. \ref{App_eq:I_oper_dfn} and \ref{App_eq:Q_oper_dfn}). The solution to eq~\ref{eq:q_A1*_A0} can be written in terms of the eigenfunctions of $\mathscr{L}_{-1}$, $\pmb{\hat{q}_{-1,k}}$, and their adjoints as follows,
\begin{equation}
\label{eq:q_A1*_A0_eig_expa}
    \pmb{\hat{q}_{A_1^{*} A_0}}
    =
    \mathlarger{\sum}_k
    \Bigg[
     \frac
    {
    \Big\langle
    \pmb{\hat{q}^{\dagger}_{-1,k}}
    ,
    i
    \Big(
    \mathbb{I}_{(0,-1)} \{ \pmb{q_0}, \pmb{\hat{q}_0} \}
    \pmb{\hat{q}_1^{*}}
    +
    \mathbb{I}_{(-1,0)} \{ \pmb{q_0}, \pmb{\hat{q}_1^{*}} \}
    \pmb{\hat{q}_0}
    +
    \mathbb{Q}_0 \{ \pmb{q_0}, \pmb{\hat{q}_0}, \pmb{\hat{q}_1^{*}} \} \pmb{q_0}
    \Big)
    \Big\rangle
    }
    {
    (\Delta\omega + \omega_{-1,k})
    \langle 
    \pmb{\hat{q}^{\dagger}_{-1,k}}
    ,
    \mathscr{B}_0
    \pmb{\hat{q}_{-1,k}}
    \rangle
    }
    \pmb{\hat{q}_{-1,k}}
    \Bigg]
\end{equation}
where, the summation is over the entire eigenspace of $\mathscr{L}_{-1}$. Thus, eq.~\ref{eq:q_A1*_A0_eig_expa} shows that when $\Delta\omega$ is small, i.e., when the axisymmetric and helical mode natural frequencies are close, the magnitude of $\pmb{\hat{q}_{A_1^* A_0}}$ is large. Therefore, when the forcing is near-resonant, i.e., $\omega_{a}\sim\omega_{0}$, $\beta_{A_{0} A_{1}}$ becomes large when $\omega_{1} \sim \omega_{a}$, resulting in an efficient reduction in the amplitude of the axisymmetric hydrodynamic response as eq.~\ref{eq:A_TH_Sol} suggests and Fig.~\ref{fig:dlr_freqs} (right) shows for the present experiment.  
\par
Thus, the asymptotic solution for the stationary flow state, up to $\epsilon^{3}$, can be written by combining eqs.~\ref{eq:Order_eps_sol},~\ref{eq:Order_eps2_gen_sol} ~\ref{eq:A_PVC_Sol},~\ref{eq:f_PVC} and \ref{eq:A_TH_Sol} as follows, 
\begin{equation}
\label{eq:full_sol}
\begin{split}
    \pmb{\tilde{q}}(r,\theta,z,t)
    =
    \pmb{q_0}
    +
    (S-S_c) \pmb{q_\Delta}
    +
    A_{PVC}^2 \pmb{q_{A_1 A^*_1}}
    ~~~~~~~~~~~~~~~~~~~~~~~~~~~~~~~~~~~~~
    \\
    +
    \Big(
    A_{PVC}
    \pmb{\hat{q}_{1}}
    e^{i \theta}
    e^{-i \omega_{PVC} t}
    +
    A_{TH}
    \pmb{\hat{q}_{0}}
    e^{-i \omega_{a} t}
    +
    c.c.
    \Big)
    ~~~~~~~~~~~~~~~~~
    \\
    +
    \Big(
    A_{PVC}^2 
    \pmb{\hat{q}_{A_1 A_1}}
    e^{2 i \theta}
    e^{-2i \omega_{PVC} t}
    +
    A_{TH}^2 \pmb{\hat{q}_{A_0 A_0}} 
    e^{-2i \omega_{a} t}
    +
    c.c.
    \Big)
    ~~~~
    \\
    +
    \Big(
    A_{PVC} A_{TH}
    \pmb{\hat{q}_{A_1 A_0}} 
    e^{i \theta} 
    e^{-i (\omega_{PVC} + \omega_{a}) t}
    +
    c.c.
    \Big)
    ~~~~~~~~~~~~~~~~~~~~
    \\
    +
    \Big(
    A_{PVC} A_{TH}
    \pmb{\hat{q}_{A_1^{*} A_0}} e^{-i \theta} 
    e^{i (\omega_{PVC} - \omega_{a}) t}
    +
    c.c.
    \Big)
    ~~~~~~~~~~~~~~~~~~~~
    \\
    +
    O\big((S-S_c)^{3/2}\big))
    ~~~~~~~~~~~~~~~~~~~~~~~~~~~~~~~~~~~~~~~~~~~~~~~~~~~~~~
\end{split}
\end{equation}
where, $t_{1}$ and $t_{2}$ have been replaced by $t$. It is now evident from eq.~\ref{eq:full_sol} that in general, the flow solution must have oscillating components at the thermoacoustic frequency, the PVC frequency, and their sum and difference. This result is consistent with the SPOD spectra for the velocity field shown in Fig.~\ref{fig:spod_velocity}. The suppression of the axisymmetric hydrodynamic response to acoustic forcing, due to the presence of a PVC, can potentially result in the suppression of the component of global heat release oscillation due to burning area oscillations in the case of an axisymmetric flame \citep{acharya2012,moeck2012PVCandAcoustics}. Therefore, ensuring that the characteristic frequency of PVC oscillations in a combustor nozzle matches those of potentially unstable thermoacoustic modes during initial stages of design might prove beneficial because the presence of a suitably designed PVC could suppress global heat release rate oscillations through the nonlinear mechanism described in this section as the qualitative agreement between flow dynamics in the present experimental study and the theory shows. 

%% file: Appendix_A.tex
The governing equations for $\pmb{\tilde{q}} = \big[ \tilde{\rho}, \tilde{u}_r, \tilde{u}_\theta, \tilde{u}_z, \tilde{p} \big]^T$ i.e. the coherent flow components are represented in the operator form as given in Eq. \ref{eq:Evol_equ_q_tilde}. The operators $\mathscr{B}$ and $\mathscr{B}^S\{\pmb{\tilde{q}}\}$ are diagonal matrices with elements $\mathscr{B} = \begin{bmatrix} 1 ~~	0 ~~ 0 ~~ 0 ~~ 0 \end{bmatrix}$ and $\mathscr{B}^s\{\pmb{\tilde{q}}\} = \begin{bmatrix} 0 ~~ \tilde{\rho} ~~ S \tilde{\rho} ~~ \tilde{\rho} ~~ 0 \end{bmatrix}$. The operator $\mathscr{B}^S \{\pmb{\tilde{q}} \}$ can then be expanded as follows:
\begin{equation}
    \mathscr{B}^S\{\pmb{\tilde{q}}\}
    =
    \mathscr{B}^S_1\{\pmb{\tilde{q}}\}
    +
    \epsilon^2 \Delta_s \mathscr{B}^S_2\{\pmb{\tilde{q}}\}
\end{equation}
where, $\mathscr{B}^S_1\{\pmb{\tilde{q}}\}$ and $\mathscr{B}^S_2\{\pmb{\tilde{q}}\}$ are diagonal matrices with elements $\mathscr{B}^S_1\{\pmb{\tilde{q}}\} = \begin{bmatrix} 0 ~~ \tilde{\rho} ~~ S_c \tilde{\rho} ~~ \tilde{\rho} ~~ 0 \end{bmatrix}$ and $\mathscr{B}^S_2\{\pmb{\tilde{q}}\} = \begin{bmatrix} 0 ~~ 0 ~~ \tilde{\rho} ~~ 0 ~~ 0 \end{bmatrix}$. The matrix $\mathscr{B}_0$ is defined as:
\begin{equation}
\label{App_eq:B0_oper_dfn}
    \mathscr{B}_0
    =
    \mathscr{B}
    +
    \mathscr{B}^S_1 \{\pmb{q_0}\}
\end{equation}
The nonlinear operators $\mathscr{N}\{\pmb{\tilde{q}}\}$, $\mathscr{N}^{S}\{\pmb{\tilde{q}}\}$ and $\mathscr{N}^{SS}\{\pmb{\tilde{q}}\}$ in eq. \ref{eq:Evol_equ_q_tilde} acting on the vector field $\pmb{\tilde{q}}$, are defined as follows:
\\
\begin{equation}
\label{App_eq:N_oper_dfn}
	\mathscr{N}\{\pmb{\tilde{q}}\}
	=
	\mathscr{N}_1\{\pmb{\tilde{q}}\}
	+
	\mathscr{R}^{\rho}_{2,4}\{\pmb{\tilde{q}}\}
	\mathscr{N}_2\{\pmb{\tilde{q}}\}
	+
	\mathscr{R}^{\rho}_{5}\{\pmb{\tilde{q}}\}
	\mathscr{R}^{\rho}_{5}\{\pmb{\tilde{q}}\}
	\mathscr{N}_3\{\pmb{\tilde{q}}\}
\end{equation}
where,
\begin{equation}
	\mathscr{N}_1\{\pmb{\tilde{q}}\}
	=
	\begin{bmatrix}
	\tilde{\Delta}_{rz}	
	&&	
	\tilde{\rho} \left(
	\frac{1}{r}
	+
	\frac{\partial }{\partial r}
	\right)	
	&&	0	&&	
	\tilde{\rho} \frac{\partial }{\partial z}
	&&	0
	\\
	\\
	0	&&	0	&&	0	&&	0	&&	0
	\\
	\\
	0	&&	0	&&	0	&&	0	&&	0
	\\
	\\
	0	&&	0	&&	0	&&	0	&&	0
	\\
	\\
	\frac{1}{Re Pr}
	\left[
	\tilde{\rho} \Lambda_{r \theta z}
	-
	2 \left(
	\frac{\partial \tilde{\rho}}{\partial r} \frac{\partial}{\partial r}
	+
	\frac{1}{r^2}
	\frac{\partial \tilde{\rho}}{\partial \theta} \frac{\partial}{\partial \theta}
	+
	\frac{\partial \tilde{\rho}}{\partial z} \frac{\partial}{\partial z}
	\right)
	\right]
	&&	0	&&	0	&&	0	&&	0
	\end{bmatrix}
\end{equation}
\begin{equation}
    \mathscr{N}_2\{\pmb{\tilde{q}}\}
    =
    \begin{bmatrix}
    0	&&	0	&&	0	&&	0	&&	0
    \\
    \\
    0	&&	\tilde{\Delta}_{r z}	&&	0	&&	0	&&	0
    \\
    \\
    0	&&	0	&&	0	&&	0	&&	0
    \\
    \\
    0	&&	0	&&	0	&&	\tilde{\Delta}_{r z}	&&	0
    \\
    \\
    0	&&	0	&&	0	&&	0	&&	0
    \end{bmatrix}
    ~~~~~
    \mathscr{N}_3\{\pmb{\tilde{q}}\}
    =
    \begin{bmatrix}
    0	&&	0	&&	0	&&	0	&&	0
    \\
    \\
    0	&&	0	&&	0	&&	0	&&	0
    \\
    \\
    0	&&	0	&&	0	&&	0	&&	0
    \\
    \\
    0	&&	0	&&	0	&&	0	&&	0
    \\
    \\
    0	&&	\tilde{\rho} \left(\frac{\partial }{\partial r} + \frac{1}{r} \right)	&&	0	&&	\tilde{\rho}\frac{\partial }{\partial z}	&&	0
    \end{bmatrix}
    ~~~~~~~~~~
\end{equation}
\\
Similarly, we can write $\mathscr{N}^S\{\pmb{\tilde{q}}\}$ and $\mathscr{N}^{SS}\{\pmb{\tilde{q}}\}$ as follows:
\begin{equation}
\label{App_eq:NS_oper_dfn}
	\mathscr{N}^S\{\pmb{\tilde{q}}\}
	=
	\mathscr{N}^S_1\{\pmb{\tilde{q}}\}
	+
	\mathscr{R}^{\rho}_{2,3,4}
	\{\pmb{\tilde{q}}\}
	\mathscr{N}^S_2\{\pmb{\tilde{q}}\}
	+
	\mathscr{R}^{\rho}_{5}\{\pmb{\tilde{q}}\}
	\mathscr{R}^{\rho}_{5}\{\pmb{\tilde{q}}\}
	\mathscr{N}^S_3\{\pmb{\tilde{q}}\}
\end{equation}
\begin{equation}
\label{App_eq:NSS_oper_dfn}
	\mathscr{N}^{SS}\{\pmb{\tilde{q}}\}
	=
	\mathscr{R}^{\rho}_{2,3}\{\pmb{\tilde{q}}\}
	\mathscr{N}^{SS}_2\{\pmb{\tilde{q}}\}
\end{equation}
where,

\begin{equation}
    \mathscr{N}^S_1\{\pmb{\tilde{q}}\}
    =
    \begin{bmatrix}
    \frac{\tilde{u}_\theta}{r} \frac{\partial } {\partial \theta}	&&	0	&&	\frac{\tilde{\rho}}{r} \frac{\partial } {\partial \theta}	&&	0	&&	0	
    \\
    \\
    0	&&	0	&&	0	&&	0	&&	0
    \\
    \\
    0	&&	0	&&	0	&&	0	&&	0
    \\
    \\
    0	&&	0	&&	0	&&	0	&&	0
    \\
    \\
    0	&&	0	&&	0	&&	0	&&	0
    \end{bmatrix}
    ~~~~~~~~
    \mathscr{N}^S_3\{\pmb{\tilde{q}}\}
    =
    \begin{bmatrix}
    0	&&	0	&&	0	&&	0	&&	0
    \\
    \\
    0	&&	0	&&	0	&&	0	&&	0
    \\
    \\
    0	&&	0	&&	0	&&	0	&&	0
    \\
    \\
    0	&&	0	&&	0	&&	0	&&	0
    \\
    \\
    0	&&	0	&&	\frac{\tilde{\rho}}{r} \frac{\partial }{\partial \theta}	&&	0	&&	0
    \end{bmatrix}
\end{equation}
\begin{equation}
	\mathscr{N}^S_2\{\pmb{\tilde{q}}\}
	=
	\begin{bmatrix}
	0	&&	0	&&	0	&&	0	&&	0
	\\
	\\
	0	&&
	\frac{\tilde{u}_\theta}{r} \frac{\partial }{\partial \theta}
	&&	0	&&	0	&&	0
	\\
	\\
	0	
	&&	
	0
	&&
	\big(
	\tilde{\Delta}_{rz}
	+
	\frac{\tilde{u}_r}{r}
	\big)
	&&	0	&&	0
	\\
	\\
	0	&&	0	&&	0	&&
	\frac{\tilde{u}_\theta}{r} \frac{\partial }{\partial \theta}
	&&	0
	\\
	\\
	0	&&	0	&&	0	&&	0	&&	0
	\end{bmatrix}
\end{equation}
\begin{equation}
    \mathscr{N}^{SS}_2\{\pmb{\tilde{q}}\}
    =
    \begin{bmatrix}
    0	&&	0	&&	0	&&	0	&&	0
    \\
    \\
    0	&&	0	&&	-\frac{\tilde{u}_\theta}{r}	&&	0	&&	0
    \\
    \\
    0	&&	0	&&	\frac{\tilde{u}_\theta}{r} \frac{\partial }{\partial \theta}	&&	0	&&	0
    \\
    \\
    0	&&	0	&&	0	&&	0	&&	0
    \\
    \\
    0	&&	0	&&	0	&&	0	&&	0
    \end{bmatrix}
\end{equation}
\\
\\
where we have defined $\Lambda_{r \theta z} = \left(
\frac{\partial^2}{\partial r^2}
+ 
\frac{1}{r} \frac{\partial}{\partial r} 
+
\frac{1}{r^2} \frac{\partial^2}{\partial \theta^2}
+
\frac{\partial^2}{\partial z^2}
\right)
$ and $\tilde{\Delta}_{rz} = \left(
\tilde{u}_r \frac{\partial }{\partial r}
+
\tilde{u}_z \frac{\partial }{\partial z}
\right)$.
Also, we define matrix $\mathscr{R}^{\rho}_{i,j,...}\{\pmb{\tilde{q}}\}$ as a diagonal matrix with non-zero entries as $\tilde{\rho}$ on the $i^{th},j^{th},...$ indices. For example, $\mathscr{R}^{\rho}_{2,4}\{\pmb{\tilde{q}}\}$ is a diagonal matrix with elements: $\begin{bmatrix} 0	&&	\tilde{\rho}	&&	0	&&	\tilde{\rho}	&&	0 \end{bmatrix}$.
\\
\\
The nonlinear operators can be expanded in powers of $\epsilon$ as follows:
\begin{equation}
\label{App_eq:N_oper_expans}
    \mathscr{N}\{\pmb{\tilde{q}}\}
    =
    \mathbb{N}^1 
    \{ \pmb{q_0} \}
    +
    \epsilon
    \mathbb{N}^{\epsilon} 
    \{ \pmb{q_0}, \pmb{q_1} \}
    +
    \epsilon^2
    \mathbb{N}^{\epsilon^2}
    \{ \pmb{q_0}, \pmb{q_1},
    \pmb{q_2} \}
    +
    \epsilon^3
    \mathbb{N}^{\epsilon^3}
    \{ \pmb{q_0}, \pmb{q_1},
    \pmb{q_2}, \pmb{q_3} \}
    +
    ...
\end{equation}
where, 
\begin{equation}
\label{App_eq:N1_oper_dfn}
    \mathbb{N}^1 
    \{ \pmb{q_0} \}
    =
    \mathscr{N}_1\{\pmb{q_0}\}
	+
	\mathscr{R}^{\rho}_{2,4}\{\pmb{q_0}\}
	\mathscr{N}_2\{\pmb{q_0}\}
	+
	\mathscr{R}^{\rho}_{5}\{\pmb{q_0}\}
	\mathscr{R}^{\rho}_{5}\{\pmb{q_0}\}
	\mathscr{N}_3\{\pmb{q_0}\}
\end{equation}

\begin{equation}
\label{App_eq:Neps_oper_dfn}
\begin{split}
    \mathbb{N}^{\epsilon}
    \{ \pmb{q_0}, \pmb{q_1} \}
    =
    \mathscr{N}_1\{\pmb{q_1}\}
	+
	\mathscr{R}^{\rho}_{2,4}\{\pmb{q_1}\}
	\mathscr{N}_2\{\pmb{q_0}\}
	+
	\mathscr{R}^{\rho}_{2,4}\{\pmb{q_0}\}
	\mathscr{N}_2\{\pmb{q_1}\}
	~~~~~~~
	\\
	+
	2 \mathscr{R}^{\rho}_{5}\{\pmb{q_0}\}
	\mathscr{R}^{\rho}_{5}\{\pmb{q_1}\}
	\mathscr{N}_3\{\pmb{q_0}\}
	+
	\mathscr{R}^{\rho}_{5}\{\pmb{q_0}\}
	\mathscr{R}^{\rho}_{5}\{\pmb{q_0}\}
	\mathscr{N}_3\{\pmb{q_1}\}
\end{split}
\end{equation}

\begin{equation}
\label{App_eq:Neps2_oper_dfn}
\begin{split}
    \mathbb{N}^{\epsilon^2}
    \{ \pmb{q_0}, \pmb{q_1}, \pmb{q_2} \}
    =
    \mathscr{N}_1\{\pmb{q_2}\}
	+
	\mathscr{R}^{\rho}_{2,4}\{\pmb{q_2}\}
	\mathscr{N}_2\{\pmb{q_0}\}
	+
	\mathscr{R}^{\rho}_{2,4}\{\pmb{q_0}\}
	\mathscr{N}_2\{\pmb{q_2}\}
    ~~~~~~~
	\\
	+
	2 \mathscr{R}^{\rho}_{5}\{\pmb{q_0}\}
	\mathscr{R}^{\rho}_{5}\{\pmb{q_2}\}
	\mathscr{N}_3\{\pmb{q_0}\}
	+
	\mathscr{R}^{\rho}_{5}\{\pmb{q_0}\}
	\mathscr{R}^{\rho}_{5}\{\pmb{q_0}\}
	\mathscr{N}_3\{\pmb{q_2}\}
	\\
	+
	\mathscr{R}^{\rho}_{2,4}\{\pmb{q_1}\}
	\mathscr{N}_2\{\pmb{q_1}\}
	+
	2 \mathscr{R}^{\rho}_{5}\{\pmb{q_0}\}
	\mathscr{R}^{\rho}_{5}\{\pmb{q_1}\}
	\mathscr{N}_3\{\pmb{q_1}\}
	~~~~~~~~
	\\
	+
	\mathscr{R}^{\rho}_{5}\{\pmb{q_1}\}
	\mathscr{R}^{\rho}_{5}\{\pmb{q_1}\}
	\mathscr{N}_3\{\pmb{q_0}\}
	~~~~~~~~~~~~~~~~~~~~~~~~~~~~~~~~~~
\end{split}
\end{equation}

\begin{equation}
\label{App_eq:Neps3_oper_dfn}
\begin{split}
    \mathbb{N}^{\epsilon^3}
    \{ \pmb{q_0}, \pmb{q_1}, \pmb{q_2}, \pmb{q_3} \}
    =
    \mathscr{N}_1\{\pmb{q_3}\}
	+
	\mathscr{R}^{\rho}_{2,4}\{\pmb{q_3}\}
	\mathscr{N}_2\{\pmb{q_0}\}
	+
	\mathscr{R}^{\rho}_{2,4}\{\pmb{q_0}\}
	\mathscr{N}_2\{\pmb{q_3}\}
	~~~~~~~~~
	\\
	+
	2 \mathscr{R}^{\rho}_{5}\{\pmb{q_0}\}
	\mathscr{R}^{\rho}_{5}\{\pmb{q_3}\}
	\mathscr{N}_3\{\pmb{q_0}\}
	+
	\mathscr{R}^{\rho}_{5}\{\pmb{q_0}\}
	\mathscr{R}^{\rho}_{5}\{\pmb{q_0}\}
	\mathscr{N}_3\{\pmb{q_3}\}
	~~
	\\
	+
	\mathscr{R}^{\rho}_{2,4}\{\pmb{q_1}\}
	\mathscr{N}_2\{\pmb{q_2}\}
	+
	2 \mathscr{R}^{\rho}_{5}\{\pmb{q_0}\}
	\mathscr{R}^{\rho}_{5}\{\pmb{q_1}\}
	\mathscr{N}_3\{\pmb{q_2}\}
	~~~~~~~~~~
	\\
	+
	\mathscr{R}^{\rho}_{5}\{\pmb{q_1}\}
	\mathscr{R}^{\rho}_{5}\{\pmb{q_2}\}
	\mathscr{N}_3\{\pmb{q_0}\}
	+
	\mathscr{R}^{\rho}_{2,4}\{\pmb{q_2}\}
	\mathscr{N}_2\{\pmb{q_1}\}
	~~~~~~~~~~~~
	\\
	+
	2 \mathscr{R}^{\rho}_{5}\{\pmb{q_0}\}
	\mathscr{R}^{\rho}_{5}\{\pmb{q_2}\}
	\mathscr{N}_3\{\pmb{q_1}\}
	+
	\mathscr{R}^{\rho}_{5}\{\pmb{q_2}\}
	\mathscr{R}^{\rho}_{5}\{\pmb{q_1}\}
	\mathscr{N}_3\{\pmb{q_0}\}
	~~
	\\
	+
	\mathscr{R}^{\rho}_{5}\{\pmb{q_1}\}
	\mathscr{R}^{\rho}_{5}\{\pmb{q_1}\}
	\mathscr{N}_3\{\pmb{q_1}\}
	~~~~~~~~~~~~~~~~~~~~~~~~~~~~~~~~~~~~
\end{split}
\end{equation}
\\
Similarly,
\begin{equation}
\label{App_eq:NS_oper_expans}
    \mathscr{N}^S\{\pmb{\tilde{q}}\}
    =
    \mathbb{N}^{S,1}
    \{ \pmb{q_0} \}
    +
    \epsilon
    \mathbb{N}^{S,\epsilon} 
    \{ \pmb{q_0}, \pmb{q_1} \}
    +
    \epsilon^2
    \mathbb{N}^{S,\epsilon^2}
    \{ \pmb{q_0}, \pmb{q_1},
    \pmb{q_2} \}
    +
    \epsilon^3
    \mathbb{N}^{S,\epsilon^3}
    \{ \pmb{q_0}, \pmb{q_1},
    \pmb{q_2}, \pmb{q_3} \}
    +
    ...
\end{equation}
where,
\begin{equation}
\label{App_eq:NS1_oper_dfn}
    \mathbb{N}^{S,1}
    \{ \pmb{q_0} \}
    =
    \mathscr{N}^S_1\{\pmb{q_0}\}
	+
	\mathscr{R}^{\rho}_{2,3,4}\{\pmb{q_0}\}
	\mathscr{N}^S_2\{\pmb{q_0}\}
	+
	\mathscr{R}^{\rho}_{5}\{\pmb{q_0}\}
	\mathscr{R}^{\rho}_{5}\{\pmb{q_0}\}
	\mathscr{N}^S_3\{\pmb{q_0}\}
\end{equation}

\begin{equation}
\label{App_eq:NSeps_oper_dfn}
\begin{split}
    \mathbb{N}^{S,\epsilon}
    \{ \pmb{q_0}, \pmb{q_1} \}
    =
    \mathscr{N}^S_1\{\pmb{q_1}\}
	+
	\mathscr{R}^{\rho}_{2,3,4}\{\pmb{q_1}\}
	\mathscr{N}^S_2\{\pmb{q_0}\}
	+
	\mathscr{R}^{\rho}_{2,3,4}\{\pmb{q_0}\}
	\mathscr{N}^S_2\{\pmb{q_1}\}
	~~
	\\
	+
	2 \mathscr{R}^{\rho}_{5}\{\pmb{q_0}\}
	\mathscr{R}^{\rho}_{5}\{\pmb{q_1}\}
	\mathscr{N}^S_3\{\pmb{q_0}\}
	+
	\mathscr{R}^{\rho}_{5}\{\pmb{q_0}\}
	\mathscr{R}^{\rho}_{5}\{\pmb{q_0}\}
	\mathscr{N}^S_3\{\pmb{q_1}\}
\end{split}
\end{equation}

\begin{equation}
\label{App_eq:NSeps2_oper_dfn}
\begin{split}
    \mathbb{N}^{S,\epsilon^2}
    \{ \pmb{q_0}, \pmb{q_1}, \pmb{q_2} \}
    =
    \mathscr{N}^S_1\{\pmb{q_2}\}
	+
	\mathscr{R}^{\rho}_{2,3,4}\{\pmb{q_2}\}
	\mathscr{N}^S_2\{\pmb{q_0}\}
	+
	\mathscr{R}^{\rho}_{2,3,4}\{\pmb{q_0}\}
	\mathscr{N}^S_2\{\pmb{q_2}\}
	~~
	\\
	+
	2 \mathscr{R}^{\rho}_{5}\{\pmb{q_0}\}
	\mathscr{R}^{\rho}_{5}\{\pmb{q_2}\}
	\mathscr{N}^S_3\{\pmb{q_0}\}
	+
	\mathscr{R}^{\rho}_{5}\{\pmb{q_0}\}
	\mathscr{R}^{\rho}_{5}\{\pmb{q_0}\}
	\mathscr{N}^S_3\{\pmb{q_2}\}
	\\
	+
	\mathscr{R}^{\rho}_{2,3,4}\{\pmb{q_1}\}
	\mathscr{N}^S_2\{\pmb{q_1}\}
	+
	2 \mathscr{R}^{\rho}_{5}\{\pmb{q_0}\}
	\mathscr{R}^{\rho}_{5}\{\pmb{q_1}\}
	\mathscr{N}^S_3\{\pmb{q_1}\}
	~~~~~~
	\\
	+
	\mathscr{R}^{\rho}_{5}\{\pmb{q_1}\}
	\mathscr{R}^{\rho}_{5}\{\pmb{q_1}\}
	\mathscr{N}^S_3\{\pmb{q_0}\}
	~~~~~~~~~~~~~~~~~~~~~~~~~~~~~~~~~~~~
\end{split}
\end{equation}

\begin{equation}
\label{App_eq:NSeps3_oper_dfn}
\begin{split}
    \mathbb{N}^{S,\epsilon^3}
    \{ \pmb{q_0}, \pmb{q_1}, \pmb{q_2}, \pmb{q_3} \}
    =
    \mathscr{N}^S_1\{\pmb{q_3}\}
	+
	\mathscr{R}^{\rho}_{2,3,4}\{\pmb{q_3}\}
	\mathscr{N}^S_2\{\pmb{q_0}\}
	+
	\mathscr{R}^{\rho}_{2,3,4}\{\pmb{q_0}\}
	\mathscr{N}^S_2\{\pmb{q_3}\}
	~~~~
	\\
	+
	2 \mathscr{R}^{\rho}_{5}\{\pmb{q_0}\}
	\mathscr{R}^{\rho}_{5}\{\pmb{q_3}\}
	\mathscr{N}^S_3\{\pmb{q_0}\}
	+
	\mathscr{R}^{\rho}_{5}\{\pmb{q_0}\}
	\mathscr{R}^{\rho}_{5}\{\pmb{q_0}\}
	\mathscr{N}^S_3\{\pmb{q_3}\}
	~~
	\\
	+
	\mathscr{R}^{\rho}_{2,3,4}\{\pmb{q_1}\}
	\mathscr{N}^S_2\{\pmb{q_2}\}
	+
	2 \mathscr{R}^{\rho}_{5}\{\pmb{q_0}\}
	\mathscr{R}^{\rho}_{5}\{\pmb{q_1}\}
	\mathscr{N}^S_3\{\pmb{q_2}\}
	~~~~~~~~
	\\
	+
	\mathscr{R}^{\rho}_{5}\{\pmb{q_1}\}
	\mathscr{R}^{\rho}_{5}\{\pmb{q_2}\}
	\mathscr{N}^S_3\{\pmb{q_0}\}
	+
	\mathscr{R}^{\rho}_{2,3,4}\{\pmb{q_2}\}
	\mathscr{N}^S_2\{\pmb{q_1}\}
	~~~~~~~~~
	\\
	+
	2 \mathscr{R}^{\rho}_{5}\{\pmb{q_0}\}
	\mathscr{R}^{\rho}_{5}\{\pmb{q_2}\}
	\mathscr{N}^S_3\{\pmb{q_1}\}
	+
	\mathscr{R}^{\rho}_{5}\{\pmb{q_2}\}
	\mathscr{R}^{\rho}_{5}\{\pmb{q_1}\}
	\mathscr{N}^S_3\{\pmb{q_0}\}
	~
	\\
	+
	\mathscr{R}^{\rho}_{5}\{\pmb{q_1}\}
	\mathscr{R}^{\rho}_{5}\{\pmb{q_1}\}
	\mathscr{N}^S_3\{\pmb{q_1}\}
	~~~~~~~~~~~~~~~~~~~~~~~~~~~~~~~~~~~~~
\end{split}
\end{equation}
\\
$\mathscr{N}^{SS}\{\pmb{\tilde{q}}\}$ is also expanded in similar way:
\begin{equation}
\label{App_eq:NSS_oper_expans}
    \mathscr{N}^{SS}\{\pmb{\tilde{q}}\}
    =
    \mathbb{N}^{SS,1}
    \{ \pmb{q_0} \}
    +
    \epsilon
    \mathbb{N}^{SS,\epsilon}
    \{ \pmb{q_0}, \pmb{q_1} \}
    +
    \epsilon^2
    \mathbb{N}^{SS,\epsilon^2}
    \{ \pmb{q_0}, \pmb{q_1},
    \pmb{q_2} \}
    +
    \epsilon^3
    \mathbb{N}^{SS,\epsilon^3}
    \{ \pmb{q_0}, \pmb{q_1},
    \pmb{q_2}, \pmb{q_3} \}
    +
    ...
\end{equation}
where,
\begin{equation}
\label{App_eq:NSS1_oper_dfn}
\begin{split}
    \mathbb{N}^{SS,1}
    \{ \pmb{q_0} \}
    =
	\mathscr{R}^{\rho}_{2,3}
	\{\pmb{q_0}\}
	\mathscr{N}^{SS}_2
	\{\pmb{q_0}\}
\end{split}
\end{equation}

\begin{equation}
\label{App_eq:NSSeps_oper_dfn}
\begin{split}
    \mathbb{N}^{SS,\epsilon}
    \{ \pmb{q_0}, \pmb{q_1} \}
    =
    \mathscr{R}^{\rho}_{2,3}\{\pmb{q_0}\}
	\mathscr{N}^{SS}_2\{\pmb{q_1}\}
	+
	\mathscr{R}^{\rho}_{2,3}\{\pmb{q_1}\}
	\mathscr{N}^{SS}_2\{\pmb{q_0}\}
\end{split}
\end{equation}

\begin{equation}
\label{App_eq:NSSeps2_oper_dfn}
\begin{split}
    \mathbb{N}^{SS,\epsilon^2}
    \{ \pmb{q_0}, \pmb{q_1}, \pmb{q_2} \}
    =
    \mathscr{R}^{\rho}_{2,3}\{\pmb{q_0}\}
	\mathscr{N}^{SS}_2\{\pmb{q_2}\}
	+
	\mathscr{R}^{\rho}_{2,3}\{\pmb{q_2}\}
	\mathscr{N}^{SS}_2\{\pmb{q_0}\}
	+
	\mathscr{R}^{\rho}_{2,3}\{\pmb{q_1}\}
	\mathscr{N}^{SS}_2\{\pmb{q_1}\}
\end{split}
\end{equation}

\begin{equation}
\label{App_eq:NSSeps3_oper_dfn}
\begin{split}
    \mathbb{N}^{SS,\epsilon^3}
    \{ \pmb{q_0}, \pmb{q_1}, \pmb{q_2}, \pmb{q_3} \}
    =
    \mathscr{R}^{\rho}_{2,3}\{\pmb{q_0}\}
	\mathscr{N}^{SS}_2\{\pmb{q_3}\}
	+
	\mathscr{R}^{\rho}_{2,3}\{\pmb{q_3}\}
	\mathscr{N}^{SS}_2\{\pmb{q_0}\}
	~~~~~~~~~~~~~~~~
	\\
	+
	\mathscr{R}^{\rho}_{2,3}\{\pmb{q_1}\}
	\mathscr{N}^{SS}_2\{\pmb{q_2}\}
	+
	\mathscr{R}^{\rho}_{2,3}\{\pmb{q_2}\}
	\mathscr{N}^{SS}_2\{\pmb{q_1}\}
	~~~~~~~~~~~~~~
\end{split}
\end{equation}
\\
We define the linearized Navier-Stokes operator i.e. $\mathscr{L}$ acting on any vector field $\pmb{p}$ as follows:
\begin{equation}
\label{App_eq:L_oper_dfn}
\begin{split}
    \mathscr{L} \pmb{p}
    =
    \big(
    \mathbb{N}^{1}
    \{ \pmb{q_0} \}
    \pmb{p}
    +
    \mathbb{N}^{\epsilon}
    \{ \pmb{q_0}, \pmb{p} \}
    \pmb{q_0}
    \big)
    +
    S_c
    \big(
    \mathbb{N}^{S,1}
    \{ \pmb{q_0} \}
    \pmb{p}
    +
    \mathbb{N}^{S,\epsilon}
    \{ \pmb{q_0}, \pmb{p} \}
    \pmb{q_0}
    \big)
    \\
    +
    ~
    S_c^2
    \big(
    \mathbb{N}^{SS,1}
    \{ \pmb{q_0} \}
    \pmb{p}
    +
    \mathbb{N}^{SS,\epsilon}
    \{ \pmb{q_0}, \pmb{p} \}
    \pmb{q_0}
    \big)
    -
    \big(
    \mathcal{L}_T \pmb{p}
    +
    S_c \mathcal{L}_T^{S} \pmb{p}
    \big)
    ~~~~~~~
\end{split}
\end{equation}
\\
We define the vector field $\pmb{p}(r,\theta,z,t_1,t_2) = A_p(t_2) \pmb{\hat{p}_m} (r,z) e^{i(m \theta - \omega t_1)}$. The operator $\mathscr{L}_m$ is obtained from $\mathscr{L}$ by the transformation $\partial/\partial \theta \to i m$ and the operation $\mathscr{L}_m \pmb{\hat{p}_m}$ can be written as follows:
\begin{equation}
\label{App_eq:Lm_oper_dfn}
\begin{split}
    \mathscr{L}_m \pmb{\hat{p}_m}
    =
    \big(
    \mathbb{N}^{1}_{(0,m)}
    \{ \pmb{q_0} \}
    \pmb{\hat{p}_m}
    +
    \mathbb{N}^{\epsilon}_{(m,0)}
    \{ \pmb{q_0}, \pmb{\hat{p}_m} \}
    \pmb{q_0}
    \big)
    +
    S_c
    \big(
    \mathbb{N}^{S,1}_m
    \{ \pmb{q_0} \}
    \pmb{\hat{p}_m}
    +
    \mathbb{N}^{S,\epsilon}_m
    \{ \pmb{q_0}, \pmb{\hat{p}_m} \}
    \pmb{q_0}
    \big)
    \\
    +
    ~
    S_c^2
    \big(
    \mathbb{N}^{SS,1}_m
    \{ \pmb{q_0} \}
    \pmb{\hat{p}_m}
    +
    \mathbb{N}^{SS,\epsilon}_m
    \{ \pmb{q_0}, \pmb{\hat{p}_m} \}
    \pmb{q_0}
    \big)
    -
    \big(
    \mathcal{L}_{T,m} \pmb{\hat{p}_m}
    +
    S_c \mathcal{L}_{T,m}^{S} \pmb{\hat{p}_m}
    \big)
    ~~~~~~~~~~~
\end{split}
\end{equation}
\\
We now define following vector fields to explain the various operator-vector operations used in eq. \ref{App_eq:Lm_oper_dfn}:
\begin{equation}
\begin{split}
    \pmb{r}
    (r,\theta,z,t_1,t_2)
    =
    A_r(t_2)
    \pmb{\hat{r}} (r,z)
    e^{i(r \theta - \omega_r t_1)}
    \\
    \pmb{s}
    (r,\theta,z,t_1,t_2)
    =
    A_s(t_2)
    \pmb{\hat{s}} (r,z)
    e^{i(s \theta - \omega_s t_1)}
\end{split}
\end{equation}
\\
$\mathbb{N}^1 \{ \pmb{q_0} \}\pmb{s}$ operation can be expanded as follows:
\begin{equation}
    \mathbb{N}^1 
    \{ \pmb{q_0} \}
    \pmb{s}
    =
    A_s e^{i s \theta} e^{-i \omega_s t_1} 
    \mathbb{N}^{1}_{(0,s)}
    \{ \pmb{q_0} \} 
    \pmb{\hat{s}}
\end{equation}
Using eq. \ref{App_eq:N1_oper_dfn},
\begin{equation}
    \mathbb{N}^{1}_{(0,s)}
    \{ \pmb{q_0} \} 
    \pmb{\hat{s}}
    =
    \mathscr{N}_1^{(0,s)} \{ \pmb{q_0} \}
    \pmb{\hat{s}}
	+
	\mathscr{R}^{\rho}_{2,4} \{ \pmb{q_0} \}
	\mathscr{N}_2 \{ \pmb{q_0} \}
	\pmb{\hat{s}}
	+
	\mathscr{R}^{\rho}_{5} \{ \pmb{q_0} \}
	\mathscr{R}^{\rho}_{5} \{ \pmb{q_0} \}
	\mathscr{N}_3 \{ \pmb{q_0} \}
	\pmb{\hat{s}}
\end{equation}
Similarly,
\begin{equation}
    \mathbb{N}^{\epsilon} 
    \{ \pmb{q_0}, \pmb{r} \}
    \pmb{s}
    =
    A_r A_s
    e^{i(r+s)\theta}
    e^{-i (\omega_r + \omega_s) t_1 }
    \mathbb{N}^{\epsilon}_{(r,s)}
    \{ \pmb{q_0}, \pmb{\hat{r}} \}
    \pmb{\hat{s}}
\end{equation}
Using eq. \ref{App_eq:Neps_oper_dfn},
\begin{equation}
\begin{split}
    \mathbb{N}^{\epsilon}_{(r,s)}
    \{ \pmb{q_0}, \pmb{r} \}
    \pmb{s}
    =
    \mathscr{N}_1^{(r,s)}
    \{ \pmb{\hat{r}} \}
    \pmb{\hat{s}}
    +
	\mathscr{R}^{\rho}_{2,4}
	\{ \pmb{q_0} \}
	\mathscr{N}_2
	\{ \pmb{\hat{r}} \}
    \pmb{\hat{s}}
	+
	\mathscr{R}^{\rho}_{2,4}
	\{ \pmb{\hat{r}} \}
	\mathscr{N}_2
	\{ \pmb{q_0} \}
    \pmb{\hat{s}}
    \\
    +
    ~
	2
	\mathscr{R}^{\rho}_{5}
	\{ \pmb{q_0} \}
	\mathscr{R}^{\rho}_{5}
	\{ \pmb{\hat{r}} \}
	\mathscr{N}_3
	\{ \pmb{q_0} \}
    \pmb{\hat{s}}
    +
	\mathscr{R}^{\rho}_{5}
	\{ \pmb{q_0} \}
	\mathscr{R}^{\rho}_{5}
	\{ \pmb{q_0} \}
	\mathscr{N}_3
	\{ \pmb{\hat{r}} \}
    \pmb{\hat{s}}
\end{split}
\end{equation}
The matrix-vector operation $\mathscr{N}_1^{(r,s)} \{ \pmb{\hat{r}} \} \pmb{\hat{s}}$ is obtained by setting $\partial \tilde{\rho}/\partial \theta = ir\tilde{\rho}$ in the first element of last row and $\partial/\partial \theta = i s$ in the expression for $\mathscr{N}_1 \{ \pmb{\tilde{q}} \}$ in eq. \ref{App_eq:N_oper_dfn}.
\\
\\
Similarly, we expand the various other operations used in eq. \ref{App_eq:Lm_oper_dfn} using eqs. \ref{App_eq:NS1_oper_dfn}, \ref{App_eq:NSeps_oper_dfn}, \ref{App_eq:NSS1_oper_dfn} and \ref{App_eq:NSSeps_oper_dfn} as follows:
\begin{equation}
    \mathbb{N}^{S,1}_{s}
    \{ \pmb{q_0} \} \pmb{\hat{s}}
    =
    \mathscr{N}^{S,s}_1
	\{ \pmb{q_0} \}
	\pmb{\hat{s}}
    +
    \mathscr{R}^{\rho}_{5}
	\{ \pmb{\hat{p}} \}
	\mathscr{N}^{S,s}_2
	\{ \pmb{q_0} \}
	\pmb{\hat{s}}
	+
	\mathscr{R}^{\rho}_{5}
	\{ \pmb{q_0} \}
	\mathscr{R}^{\rho}_{5}
	\{ \pmb{q_0} \}
	\mathscr{N}^{S,s}_3
	\{ \pmb{q_0} \}
	\pmb{\hat{s}}
\end{equation}
\begin{equation}
\begin{split}
    \mathbb{N}^{S,\epsilon}_{s}
    \{ \pmb{q_0}, \pmb{\hat{r}} \} \pmb{\hat{s}}
    =
    \mathscr{N}^{S,s}_1
    \{ \pmb{\hat{r}} \}
    \pmb{\hat{s}}
	+
	\mathscr{R}^{\rho}_{2,3,4}
	\{ \pmb{\hat{r}} \}
	\mathscr{N}^{S,s}_2
	\{ \pmb{q_0} \}
	\pmb{\hat{s}}
	+
	\mathscr{R}^{\rho}_{2,3,4}
	\{ \pmb{q_0} \}
	\mathscr{N}^{S,s}_2
	\{ \pmb{\hat{r}} \}
	\pmb{\hat{s}}
	\\
	+
	2 \mathscr{R}^{\rho}_{5}
	\{ \pmb{q_0} \}
	\mathscr{R}^{\rho}_{5}
	\{ \pmb{\hat{r}} \}
	\mathscr{N}^{S,s}_3
	\{ \pmb{q_0} \}
	\pmb{\hat{s}}
	+
	\mathscr{R}^{\rho}_{5}
	\{ \pmb{q_0} \}
	\mathscr{R}^{\rho}_{5}
	\{ \pmb{q_0} \}
	\mathscr{N}^{S,s}_3
	\{ \pmb{\hat{r}} \}
	\pmb{\hat{s}}
\end{split}
\end{equation}

\begin{equation}
    \mathbb{N}^{SS,1}_{s}
    \{ \pmb{q_0} \} \pmb{\hat{s}}
    =
    \mathscr{R}^{\rho}_{2,3}
	\{ \pmb{q_0} \}
	\mathscr{N}^{SS,s}_2
	\{ \pmb{q_0} \}
	\pmb{\hat{s}}
\end{equation}
\begin{equation}
\begin{split}
    \mathbb{N}^{SS,\epsilon}_{s}
    \{ \pmb{q_0}, \pmb{\hat{r}} \} \pmb{\hat{s}}
    =
    \mathscr{R}^{\rho}_{2,3}
    \{ \pmb{q_0} \}
	\mathscr{N}^{SS,s}_2
	\{ \pmb{\hat{r}} \}
	\pmb{\hat{s}}
	+
	\mathscr{R}^{\rho}_{2,3}
	\{ \pmb{\hat{r}} \}
	\mathscr{N}^{SS,s}_2
	\{ \pmb{q_0} \}
	\pmb{\hat{s}}
\end{split}
\end{equation}

%% file: Appendix_B.tex
The coefficients in the evolution equation for $A_1(t_2)$ i.e. eq. \ref{eq:A1_evol} are defined as follows:
\begin{equation}
\label{App_eq:alpha_A1}
\begin{split}
    \alpha_{A_1}
    =
    \frac
    {1}
    {
    \langle
    \pmb{\hat{q}_1^{\dagger}}, \mathscr{B}_0 \pmb{\hat{q}_1}
    \rangle
    }
    \bigg\langle
    \pmb{\hat{q}_1^{\dagger}}
    ,
    -\Big[
    (-i \omega_1)
    \mathscr{B}_1^S \{ \pmb{\hat{q}_{\Delta}} \}
    \pmb{\hat{q}_{1}}
    +
    \mathbb{S}_{(0,1)} \{ \pmb{q_0},  \pmb{\hat{q}_{\Delta}} \}
    \pmb{\hat{q}_{1}}
    +
    \mathbb{S}_{(1,0)} \{ \pmb{q_0}, \pmb{\hat{q}_{1}} \} \pmb{\hat{q}_{\Delta}}
    \\
    +
    ~
    \mathbb{R}_{0} \{ \pmb{q_0}, \pmb{\hat{q}_{1}}, \pmb{\hat{q}_{\Delta}} \}
    \pmb{q_0}
    +
    \mathbb{R}_{0} \{ \pmb{q_0}, \pmb{\hat{q}_{\Delta}}, \pmb{\hat{q}_{1}} \}
    \pmb{q_0}
    +
    \mathbb{N}_{1}^{S,1}
    \{ \pmb{q_0} \}
    \pmb{\hat{q}_{1}}
    ~~~~~~~
    \\
    +
    ~
    2 S_c
    \mathbb{N}_{1}^{SS,1}
    \{ \pmb{q_0} \}
    \pmb{\hat{q}_{1}}
    -
    \mathcal{L}_{T,0}^S \pmb{\hat{q}_{1}}
    +
    (-i \omega_1)
    \mathscr{B}^S_2 \{ \pmb{q_0} \}
    \pmb{\hat{q}_{1}}
    ~~~~~~~~~~~~
    \\
    +
    ~
    \mathbb{N}^{S,\epsilon}_{0}
    \{ \pmb{q_0}, \pmb{\hat{q}_{1}} \}
    \pmb{q_0}
    +
    2 S_c 
    \mathbb{N}^{SS,\epsilon}_{0}
    \{ \pmb{q_0}, \pmb{\hat{q}_{1}} \}
    \pmb{q_0}
    \Big]
    \bigg\rangle
    ~~~~~~~~~~~~~~~~~~~~
\end{split}
\end{equation}
\\
\begin{equation}
\label{App_eq:beta_A1_A1}
\begin{split}
    \beta_{A_1 A_1}
    =
    \frac
    {1}
    {
    \langle
    \pmb{\hat{q}_1^{\dagger}}, \mathscr{B}_0 \pmb{\hat{q}_1}
    \rangle
    }
    \bigg\langle
    \pmb{\hat{q}_1^{\dagger}}
    ,
    \Big[
    \mathbb{I}_{(-1,2)} \{ \pmb{q_0}, \pmb{\hat{q}_1^{*}} \}
    \pmb{\hat{q}_{A_1 A_1}}
    +
    \mathbb{I}_{(0,1)} \{ \pmb{q_0}, \pmb{\hat{q}_{A_1 A_1^{*}}} \}
    \pmb{\hat{q}_{1}}
    +
    \mathbb{I}_{(2,-1)} \{ \pmb{q_0}, \pmb{\hat{q}_{A_1 A_1}} \}
    \pmb{\hat{q}_1}
    \\
    +
    ~
    \mathbb{S}_{(1,0)} \{ \pmb{q_0}, \pmb{\hat{q}_{1}} \} \pmb{\hat{q}_{A_1 A_1^{*}}}
    +
    \mathbb{R}_{-1} \{ \pmb{q_0}, \pmb{\hat{q}_{1}}, \pmb{\hat{q}_{1}} \}
    \pmb{\hat{q}_1^{*}}
    +
    \mathbb{Q}_{1} \{ \pmb{q_0}, \pmb{\hat{q}_1^{*}}, \pmb{\hat{q}_{1}} \}
    \pmb{\hat{q}_{1}}
    ~~~~~~~~
    \\
    +
    ~
    \mathbb{Q}_{0} \{ \pmb{q_0}, \pmb{\hat{q}_{1}}, \pmb{\hat{q}_{A_1 A_1^{*}}} \}
    \pmb{q_0}
    +
    \mathbb{Q}_{0} \{ \pmb{q_0}, \pmb{\hat{q}_1^{*}}, \pmb{\hat{q}_{A_1 A_1}} \}
    \pmb{q_0}
    +
    \mathbb{P}^{\epsilon^3}
    \{ \pmb{\hat{q}_{1}}, \pmb{\hat{q}_{1}}, \pmb{\hat{q}_1^{*}} \}
    \pmb{q_0}
    ~~~~
    \\
    +
    ~
    \mathbb{P}^{\epsilon^3}
    \{ \pmb{\hat{q}_{1}}, \pmb{\hat{q}_1^{*}}, \pmb{\hat{q}_{1}} \}
    \pmb{q_0}
    +
    \mathbb{P}^{\epsilon^3}
    \{ \pmb{\hat{q}_1^{*}}, \pmb{\hat{q}_{1}}, \pmb{\hat{q}_{1}} \}
    \pmb{q_0}
    +
    S_c
    \mathbb{P}_{0}^{S,\epsilon^3}
    \{ \pmb{\hat{q}_{1}}, \pmb{\hat{q}_{1}}, \pmb{\hat{q}_1^{*}} \}
    \pmb{q_0}
    ~~~~~~~
    \\
    +
    ~
    S_c
    \mathbb{P}_{0}^{S,\epsilon^3}
    \{ \pmb{\hat{q}_{1}}, \pmb{\hat{q}_1^{*}}, \pmb{\hat{q}_{1}} \}
    \pmb{q_0}
    +
    S_c
    \mathbb{P}_{0}^{S,\epsilon^3}
    \{ \pmb{\hat{q}_1^{*}}, \pmb{\hat{q}_{1}}, \pmb{\hat{q}_{1}} \}
    \pmb{q_0}
    \Big]
    \bigg\rangle
    ~~~~~~~~~~~~~~~~~~~~~~
\end{split}
\end{equation}
\\
The coefficients in the evolution equation for $A_0(t_2)$ i.e. eq. \ref{eq:A0_evol} are defined as follows:
\\
\begin{equation}
\label{App_eq:alpha_A0}
\begin{split}
    \alpha_{A_0}
    =
    \frac
    {1}
    {
    \langle
    \pmb{\hat{q}_0^{\dagger}}, \mathscr{B}_0 \pmb{\hat{q}_0}
    \rangle
    }
    \bigg\langle
    \pmb{\hat{q}_0^{\dagger}}
    ,
    -
    \Big[
    (-i \omega_0)
    \mathscr{B}_1^S \{ \pmb{\hat{q}_{\Delta}} \}
    \pmb{\hat{q}_{0}}
    +
    \mathbb{S}_{(0,0)} \{ \pmb{q_0}, \pmb{\hat{q}_{\Delta}} \}
    \pmb{\hat{q}_{0}}
    +
    \mathbb{S}_{(0,0)} \{ \pmb{q_0}, \pmb{\hat{q}_{0}} \}
    \pmb{\hat{q}_{\Delta}}
    \\
    +
    ~
    \mathbb{R}_{0} \{ \pmb{q_0}, \pmb{\hat{q}_{0}}, \pmb{\hat{q}_{\Delta}} \}
    \pmb{q_0}
    +
    \mathbb{R}_{0} \{ \pmb{q_0}, \pmb{\hat{q}_{\Delta}}, \pmb{\hat{q}_{0}},  \}
    \pmb{q_0}
    +
    \mathbb{N}_{1}^{S,1}
    \{ \pmb{q_0} \}
    \pmb{\hat{q}_{0}}
    ~~~~~~
    \\
    +
    ~
    2 S_c
    \mathbb{N}_{1}^{SS,1}
    \{ \pmb{q_0} \}
    \pmb{\hat{q}_{0}}
    -
    \mathcal{L}_{T,0}^S \pmb{\hat{q}^{0}}
    +
    (-i \omega_0)
    \mathscr{B}^S_2 \{ \pmb{q_0} \}
    \pmb{\hat{q}_{0}}
    ~~~~~~~~~~~~
    \\
    +
    ~
    \mathbb{N}^{S,\epsilon}_{0}
    \{ \pmb{q_0}, \pmb{\hat{q}_{0}} \}
    \pmb{q_0}
    +
    2 S_c 
    \mathbb{N}^{SS,\epsilon}_{0}
    \{ \pmb{q_0}, \pmb{\hat{q}_{0}} \}
    \pmb{q_0}
    \Big]
    \bigg\rangle
    ~~~~~~~~~~~~~~~~~~~~~
\end{split}
\end{equation}
\\
\begin{equation}
\label{App_eq:Beta_A0_f}
    \beta_{A_0 f} 
    = 
    \frac
    {\langle \pmb{\hat{q}_0^{\dagger}}, \pmb{\hat{q}_a} \rangle}
    {2 \langle \pmb{\hat{q}_0^{\dagger}}, \mathscr{B}_0 \pmb{\hat{q}_{0}} \rangle}
\end{equation}
\\
\begin{equation}
\begin{split}
\label{App_eq:Beta_A0_A1}
    \beta_{A_0 A_1}
    =
    \frac{1}
    {
    \langle 
    \pmb{\hat{q}_0^{\dagger}}
    ,
    \mathscr{B}_0
    \pmb{\hat{q}_{0}}
    \rangle
    }
    \bigg\langle
    \pmb{\hat{q}^{0\dagger}}
    ,
    \Big[
    \mathbb{I}_{(-1,1)} \{ \pmb{q_0}, \pmb{\hat{q}_1^{*}} \}
    \pmb{\hat{q}_{A_1 A_0}}
    +
    \mathbb{I}_{(1,-1)} \{ \pmb{q_0}, \pmb{\hat{q}_{A_1 A_0}} \}
    \pmb{\hat{q}_1^{*}}
    +
    \mathbb{I}_{(-1,1)} \{ \pmb{q_0}, \pmb{\hat{q}_{A_1^{*} A_0}} \}
    \pmb{\hat{q}_{1}}
    \\
    +
    ~
    \mathbb{I}_{(1,-1)} \{ \pmb{q_0}, \pmb{\hat{q}_{1}} \}
    \pmb{\hat{q}_{A_1^{*} A_0}}
    +
    \mathbb{I}_{(0,0)} \{ \pmb{q_0}, \pmb{\hat{q}_{A_1 A_1^{*}}} \}
    \pmb{\hat{q}_{0}}
    +
    \mathbb{S}_{(0,0)} \{ \pmb{q_0}, \pmb{\hat{q}_{0}} \}
    \pmb{\hat{q}_{A_1 A_1^{*}}}
    ~~~~~~~~~
    \\
    +
    ~
    \mathbb{Q}_{0} \{ \pmb{q_0}, \pmb{\hat{q}_1^{*}}, \pmb{\hat{q}_{1}} \}
    \pmb{\hat{q}_{0}}
    +
    \mathbb{Q}_{1} \{ \pmb{q_0}, \pmb{\hat{q}_1^{*}}, \pmb{\hat{q}_{0}} \}
    \pmb{\hat{q}_{1}}
    +
    \mathbb{Q}_{-1} \{ \pmb{q_0}, \pmb{\hat{q}_{1}}, \pmb{\hat{q}_{0}} \}
    \pmb{\hat{q}_1^{*}}
    ~~~~~~~~~~~~~~~~~~~~
    \\
    +
    \mathbb{T}_0
    \{
    \pmb{q_{0}}, 
    \pmb{\hat{q}_{0}},
    \pmb{\hat{q}_{1}},
    \pmb{\hat{q}_1^*},
    \pmb{\hat{q}_{A_1 A_1^*}},
    \pmb{\hat{q}_{A_1 A_0}},
    \pmb{\hat{q}_{A_1^* A_0}}
    \}
    \pmb{q_{0}}
    \Big]
    \bigg\rangle
    ~~~~~~~~~~~~~~~~~~~~~~~~~~~~~~~~~~~~~
\end{split}
\end{equation}

%% file: Appendix_C.tex
We define following vector fields for describing the various matrix operators used in this paper:
\begin{equation}
\begin{split}
    \pmb{\tilde{p}} (r,\theta,z,t_1,t_2)
    =
    A_p (t_2)
    \pmb{\hat{p}} (r,z)
    e^{i(p \theta - \omega_p t_1)}
    \\
    \pmb{\tilde{r}} (r,\theta,z,t_1,t_2)
    =
    A_r (t_2)
    \pmb{\hat{r}} (r,z)
    e^{i(r \theta - \omega_r t_1)}
    ~
    \\
    \pmb{\tilde{s}} (r,\theta,z,t_1,t_2)
    =
    A_s (t_2)
    \pmb{\hat{s}} (r,z)
    e^{i(s \theta - \omega_s t_1)}
    ~
\end{split}
\end{equation}
\\
\begin{equation}
\label{App_eq:T0_q0_oper_dfn}
\begin{split}
    \mathbb{T}_0
    \{
    \pmb{q_{0}}, 
    \pmb{\hat{q}_{0}},
    \pmb{\hat{q}_{1}},
    \pmb{\hat{q}_1^*},
    \pmb{\hat{q}_{A_1 A_1^*}},
    \pmb{\hat{q}_{A_1 A_0}},
    \pmb{\hat{q}_{A_1^* A_0}}
    \}
    \pmb{q_{0}}
    =
    \Big[
    \mathbb{Q}_{0} \{ \pmb{q_0}, \pmb{\hat{q}_{0}}, \pmb{\hat{q}_{A_1 A_1^{*}}} \}
    ~~~~~~~~~~~~~~~~~~~~~~~~~~~~~~~~~
    \\
    +
    ~
    \mathbb{Q}_{0} \{ \pmb{q_0}, \pmb{\hat{q}_{1}}, \pmb{\hat{q}_{A_1^{*} A_0}} \}
    +
    \mathbb{Q}_{0} \{ \pmb{q_0}, \pmb{\hat{q}_1^*}, \pmb{\hat{q}_{A_1 A_0}} \}
    ~~~~~~~~~
    \\
    +
    ~
    \mathbb{P}^{\epsilon^3}
    \{ \pmb{\hat{q}^{0}}, \pmb{\hat{q}_{1}}, \pmb{\hat{q}_1^{*}} \}
    +
    \mathbb{P}^{\epsilon^3}
    \{ \pmb{\hat{q}_{0}}, \pmb{\hat{q}_1^{*}}, \pmb{\hat{q}_{1}} \}
    ~~~~~~~~~~~~~~~~~
    \\
    +
    ~
    \mathbb{P}^{\epsilon^3}
    \{ \pmb{\hat{q}_{1}}, \pmb{\hat{q}_{0}}, \pmb{\hat{q}_1^{*}} \}
    +
    \mathbb{P}^{\epsilon^3}
    \{ \pmb{\hat{q}_1^{*}}, \pmb{\hat{q}_{0}}, \pmb{\hat{q}_{1}} \}
    ~~~~~~~~~~~~~~~~~
    \\
    +
    ~
    \mathbb{P}^{\epsilon^3}
    \{ \pmb{\hat{q}_{1}}, \pmb{\hat{q}_1^{*}}, \pmb{\hat{q}_{0}} \}
    +
    \mathbb{P}^{\epsilon^3}
    \{ \pmb{\hat{q}_1^{*}}, \pmb{\hat{q}_{1}}, \pmb{\hat{q}_{0}} \}
    ~~~~~~~~~~~~~~~~~
    \\
    +
    ~
    S_c
    \big(
    \mathbb{P}_0^{S,\epsilon^3}
    \{ \pmb{\hat{q}_0}, \pmb{\hat{q}_1}, \pmb{\hat{q}_1^*} \}
    +
    \mathbb{P}_0^{S,\epsilon^3}
    \{ \pmb{\hat{q}_0}, \pmb{\hat{q}_1^*}, \pmb{\hat{q}_1} \}
    \big)
    ~~~~~~~
    \\
    +
    ~
    S_c
    \big(
    \mathbb{P}_0^{S,\epsilon^3}
    \{ \pmb{\hat{q}_{1}}, \pmb{\hat{q}_{0}}, \pmb{\hat{q}_1^*} \}
    +
    \mathbb{P}_0^{S,\epsilon^3}
    \{ \pmb{\hat{q}_1^*}, \pmb{\hat{q}_0}, \pmb{\hat{q}_1} \}
    \big)
    ~~~~~~~
    \\
    +
    ~
    S_c
    \big(
    \mathbb{P}_0^{S,\epsilon^3}
    \{ \pmb{\hat{q}_1}, \pmb{\hat{q}_1^*}, \pmb{\hat{q}^{0}} \}
    +
    \mathbb{P}_0^{S,\epsilon^3}
    \{ \pmb{\hat{q}_1^*}, \pmb{\hat{q}_1}, \pmb{\hat{q}_0} \}
    \big)
    \Big]
    \pmb{q_0}
    ~~~
\end{split}
\end{equation}
\\
where,
\begin{equation}
\label{App_eq:P^eps3_oper_dfn}
\begin{split}
    \mathbb{P}^{\epsilon^3} \{ \pmb{\hat{p}}, \pmb{\hat{r}}, \pmb{\hat{s}} \} \pmb{q_0}
    =
    \mathscr{R}^{\rho}_{5}
    \{\pmb{\hat{p}}\}
	\mathscr{R}^{\rho}_{5}
	\{\pmb{\hat{r}}\}
	\mathscr{N}_3
	\{\pmb{\hat{s}}\}
	\pmb{q_0}
\end{split}
\end{equation}
\begin{equation}
\label{App_eq:P^S,eps3_oper_dfn}
\begin{split}
    \mathbb{P}^{S,\epsilon^3}_{0} \{ \pmb{\hat{p}}, \pmb{\hat{r}}, \pmb{\hat{s}} \} \pmb{q_0}
    =
    \mathscr{R}^{\rho}_{5}
    \{\pmb{\hat{p}}\}
	\mathscr{R}^{\rho}_{5}
	\{\pmb{\hat{r}}\}
	\mathscr{N}^{S,0}_3
	\{\pmb{\hat{s}}\}
	\pmb{q_0}
\end{split}
\end{equation}
\begin{equation}
\label{App_eq:Q_oper_dfn}
\begin{split}
    \mathbb{Q}_{s} \{ \pmb{q_0}, \pmb{\hat{p}}, \pmb{\hat{r}} \} \pmb{\hat{s}}
    =
    \mathbb{R}_{s} \{ \pmb{q_0}, \pmb{\hat{p}}, \pmb{\hat{r}} \} \pmb{\hat{s}}
    +
    \mathbb{R}_{s} \{ \pmb{q_0}, \pmb{\hat{r}}, \pmb{\hat{p}} \} \pmb{\hat{s}}
\end{split}
\end{equation}
\begin{equation}
\label{App_eq:I_oper_dfn}
\begin{split}
    \mathbb{I}_{(p,r)} \{ \pmb{q_0}, \pmb{\hat{p}} \}
    \pmb{\hat{r}}
    =
    (-i \omega_r) \mathscr{B}^S_1 \{ \pmb{\hat{p}} \} \pmb{\hat{r}}
    +
    \mathbb{S}_{(p,r)} \{ \pmb{q_0}, \pmb{\hat{p}} \} \pmb{\hat{r}}
\end{split}
\end{equation}
\\
\begin{equation}
\label{App_eq:R_oper_dfn}
\begin{split}
    \mathbb{R}_{s} \{ \pmb{q_0}, \pmb{\hat{p}}, \pmb{\hat{r}} \}
    \pmb{\hat{s}}
    =
    \mathbb{P}^{\epsilon^2}
    \{ \pmb{q_0}, \pmb{\hat{p}}, \pmb{\hat{r}} \}
    \pmb{\hat{s}}
    +
    S_c
    \mathbb{P}^{S,\epsilon^2}_{s}
    \{ \pmb{q_0}, \pmb{\hat{p}}, \pmb{\hat{r}} \}
    \pmb{\hat{s}}
    +
    S^2_c
    \mathbb{P}^{SS,\epsilon^2}_{s}
    \{ \pmb{\hat{p}}, \pmb{\hat{r}} \}
    \pmb{\hat{s}}
\end{split}
\end{equation}
\begin{equation}
\label{App_eq:S_oper_dfn}
\begin{split}
    \mathbb{S}_{(p,r)} \{ \pmb{q_0}, \pmb{\hat{p}} \}
    \pmb{\hat{r}}
    =
    \mathbb{N}^{\epsilon}_{(p,r)}
    \{ \pmb{q_0}, \pmb{\hat{p}} \}
    \pmb{\hat{r}}
    +
    S_c
    \mathbb{N}^{S,\epsilon}_{r}
    \{ \pmb{q_0}, \pmb{\hat{r}} \}
    \pmb{\hat{r}}
    +
    S^2_c
    \mathbb{N}^{SS,\epsilon}_{r}
    \{ \pmb{q_0}, \pmb{\hat{p}} \}
    \pmb{\hat{r}}
\end{split}
\end{equation}
\\
\begin{equation}
\label{App_eq:P^eps2_oper_dfn}
\begin{split}
    \mathbb{P}^{\epsilon^2}
    \{ \pmb{q_0}, \pmb{\hat{p}}, \pmb{\hat{r}} \}
    \pmb{\hat{s}}
    =
    \mathscr{R}^{\rho}_{2,4}
    \{ \pmb{\hat{p}} \}
	\mathscr{N}_2 
	\{ \pmb{\hat{r}} \}
	\pmb{\hat{s}}
	+
    2 \mathscr{R}^{\rho}_{5}
	\{ \pmb{q_0} \}
	\mathscr{R}^{\rho}_{5}
	\{ \pmb{\hat{p}} \}
	\mathscr{N}_3
	\{ \pmb{\hat{r}} \}
	\pmb{\hat{s}}
	\\
	+
	~
	\mathscr{R}^{\rho}_{5}
	\{ \pmb{\hat{p}} \}
	\mathscr{R}^{\rho}_{5}
	\{ \pmb{\hat{r}} \}
	\mathscr{N}_3
	\{ \pmb{q_0} \}
	\pmb{\hat{s}}
	~~~~~~~~~~~~~~~~~~~~~~~~~
\end{split}
\end{equation}
\begin{equation}
\label{App_eq:P^S,eps2_oper_dfn}
\begin{split}
    \mathbb{P}^{S,\epsilon^2}_{s}
    \{ \pmb{q_0}, \pmb{\hat{p}}, \pmb{\hat{r}} \}
    \pmb{\hat{s}}
    =
    \mathscr{R}^{\rho}_{2,4}
    \{ \pmb{\hat{p}} \}
	\mathscr{N}^{S,s}_2 
	\{ \pmb{\hat{r}} \}
	\pmb{\hat{s}}
	+
    2 \mathscr{R}^{\rho}_{5}
	\{ \pmb{q_0} \}
	\mathscr{R}^{\rho}_{5}
	\{ \pmb{\hat{p}} \}
	\mathscr{N}^{S,s}_3
	\{ \pmb{\hat{r}} \}
	\pmb{\hat{s}}
	\\
	+
	~
	\mathscr{R}^{\rho}_{5}
	\{ \pmb{\hat{p}} \}
	\mathscr{R}^{\rho}_{5}
	\{ \pmb{\hat{r}} \}
	\mathscr{N}^{S,s}_3
	\{ \pmb{q_0} \}
    \pmb{\hat{s}}
    ~~~~~~~~~~~~~~~~~~~~~~~~~~~~
\end{split}
\end{equation}
\begin{equation}
\label{App_eq:P^SS,eps2_oper_dfn}
\begin{split}
    \mathbb{P}^{SS,\epsilon^2}_{s}
    \{ \pmb{\hat{p}}, \pmb{\hat{r}} \}
    \pmb{\hat{s}}
    =
    \mathscr{R}^{\rho}_{2,3}
    \{ \pmb{\hat{p}} \}
	\mathscr{N}^{SS,s}_2
	\{ \pmb{\hat{r}} \}
	\pmb{\hat{s}}
\end{split}
\end{equation}